\global\def\draftcontrol{0}

   \def\versionno{ ks  order}
\catcode`\@=11

\expandafter\ifx\csname draftcontrol\endcsname\relax\global\def\draftcontrol{0}
\fi

{\count255=\time\divide\count255 by 60
\xdef\hourmin{\number\count255}
\multiply\count255 by-60\advance\count255 by\time
\xdef\hourmin{\hourmin:\ifnum\count255<10 0\fi\the\count255}}
\def\draftdate{\number\month/\number\day/\number\year\ \ \ \hourmin }

\newcommand\makepapertitle{\par
  \begingroup
    \renewcommand\thefootnote{\@fnsymbol\c@footnote}%
    \def\@makefnmark{\rlap{\@textsuperscript{\normalfont\@thefnmark}}}%
    \long\def\@makefntext##1{\parindent 1em\noindent
            \hb@xt@1.8em{%
                \hss\@textsuperscript{\normalfont\@thefnmark}}##1}%
     \newpage
     \global\@topnum\z@   % Prevents figures from going at top of page.
     \@makepapertitle
     \thispagestyle{empty}\@thanks
  \endgroup
  \setcounter{footnote}{0}%
  \global\let\thanks\relax
  \global\let\makepapertitle\relax
  \global\let\@makepapertitle\relax
  \global\let\@thanks\@empty
  \global\let\@author\@empty
  \global\let\@date\@empty
  \global\let\@title\@empty
  \global\let\title\relax
  \global\let\author\relax
  \global\let\date\relax
  \global\let\and\relax
  \def\version{\let\version\@version\@gobble}
}
\def\@makepapertitle{%
  \newpage
   \ifnum\draftcontrol=1 {}
   \version\versionno
   \vskip 3em%
   \else
   \hfill\hbox to 3cm {\parbox{4cm}{\@pubnum}\hss}%
   \vskip 3em%
   \fi
   \begin{center}%
   \let \footnote \thanks
     {\LARGE {\@title}}%
     \vskip 1.5em%
     {\normalsize%\large
       \lineskip .5em%
       \begin{tabular}[t]{c}%
         \@author
       \end{tabular}\par}%
     \vskip 1.5em%
     {\@bstract}%
     \end{center}%
     \vskip 1.5em
     \@date%
   \par
}

\gdef\@pubnum{}
\def\pubnum#1{%
  \gdef\@pubnum{#1}}

\gdef\@bstract{}
\def\Abstract#1{%
  \gdef\@bstract{%
   \parbox{\textwidth-0pc}{%
   \centerline{\bf Abstract}\penalty1000%
\kern.2cm%
\noindent%\abstractfont \baselineskip=12pt
\renewcommand\baselinestretch{1.0}%
{#1}}}
}

\def\ps@paper{\let\@mkboth\@gobbletwo%
     \ifnum\draftcontrol=1
    \def\@oddfoot{\hbox to \textwidth{\tiny \versionno \hfil\tiny\draftdate}%
    \hskip -\textwidth \hbox to \textwidth{\hfil\rm\thepage\hfil}}%
     \else\def\@oddfoot{\hbox to \textwidth{\hfil\rm\thepage\hfil}}
     \fi
     \let\@evenfoot\@oddfoot
}
\def\body{\clearpage
          \pagestyle{paper}
    }
\def\@version#1{\ifnum\draftcontrol=1
\typeout{}\typeout{#1}\typeout{}
\vskip3mm\centerline{\hbox{\fbox{\normalsize{\tt DRAFT -- #1 -- }
                   {\draftdate}}}}\vskip3mm
\fi}
\let\version\@version
\long\def\eqlabel#1{\ifnum\draftcontrol=1
                    \tag@false  % there are some problems with multline without this
                    \tag*{(\theequation) \hbox to -0.2cm{\hspace{0cm}\small{#1}\hss}}
                    \refstepcounter{equation}
                    \edef\@currentlabel{\theequation}
                    \ltx@label{#1}          % use old LaTeX \label instead of new definition
                    \else
                    \label{#1}
                    \fi
                    }
\let\st@bibitem\@bibitem
\let\st@lbibitem\@lbibitem
\ifnum\draftcontrol=1
  \def\@bibitem#1{%
    \st@bibitem{#1}\a@@label{#1}\ignorespaces}
  \def\@lbibitem[#1]#2{%
    \st@lbibitem[#1]{#2}\a@@label{#2}\ignorespaces}
  \def\a@@label#1{%
    \gdef\a@lab{\smash{\normalfont\small#1}}
    \ifvmode
      \if@inlabel
        \global\setbox\@labels\hbox{%
          \llap{\a@lab\let\a@lab\relax
                \kern\@totalleftmargin\kern\marginparsep}%
          \box\@labels}%
      \fi
    \fi}
\fi
\documentclass[12pt,letterpaper]{article}
\usepackage{amsmath,amssymb,array,calc,epsfig,rotating,bm,xcolor}
\usepackage[sort]{cite}
\usepackage{graphicx,esint,float}
\usepackage{psfrag,verbatim}
\usepackage[makeroom]{cancel}
\usepackage{xcolor,url}
\usepackage{hyperref}
\ifnum\draftcontrol=1
\fi
\renewcommand\baselinestretch{1.25}
\renewcommand\section{\@startsection {section}{1}{\z@}%
                                   {-3.5ex \@plus -1ex \@minus -.2ex}%
                                   {2.3ex \@plus.2ex}%
                                   {\normalfont\large\bfseries}}
\renewcommand\subsection{\@startsection{subsection}{2}{\z@}%
                                   {-3.25ex\@plus -1ex \@minus -.2ex}%
                                   {1.5ex \@plus .2ex}%
                                   {\normalfont\normalsize\bfseries}}
\renewcommand\subsubsection{\@startsection{subsubsection}{3}{\z@}%
                                   {-3.25ex\@plus -1ex \@minus -.2ex}%
                                   {1.5ex \@plus .2ex}%
                                   {\normalfont\normalsize\it}}
\renewcommand\paragraph{\@startsection{paragraph}{4}{\z@}%
                                   {-3.25ex\@plus -1ex \@minus -.2ex}%
                                   {1.5ex \@plus .2ex}%
                                   {\normalfont\normalsize\bf}}

\numberwithin{equation}{section}

\def\revise#1       {\raisebox{-0em}{\rule{3pt}{1em}}%
                     \marginpar{\raisebox{.5em}{\vrule width3pt\
                     \vrule width0pt height 0pt depth0.5em
                     \hbox to 0cm{\hspace{0cm}{%
                     \parbox[t]{4em}{\raggedright\footnotesize{#1}}}\hss}}}}

\newcommand\nxt[1]  {\\\fnxt#1}
\newcommand{\ie}{{\it i.e.,}\ }

\def\calc         {{\cal C}}

\def\calf         {{\cal F}}

\def\calk         {{\cal K}}

\def\calm         {{\cal M}}
\def\caln         {{\cal N}}
\def\calo         {{\cal O}}
\def\calp         {{\cal P}}

\def\calw         {{\cal W}}

\def\reals        {{\mathbb R}}
\def\zet          {{\mathbb Z}}

\def\del          {\partial}

\def\tr           {\mathop{\rm Tr}}

\def\sqr#1#2{{\vcenter{\vbox{\hrule height.#2pt
 \hbox{\vrule width.#2pt height#1pt \kern#1pt
 \vrule width.#2pt}\hrule height.#2pt}}}}

\def\hg{\hat{g}}

\def\aa1{\phi}
\def\cc1{\psi}

\def\Om{\Omega}
\def\om{\Omega}

\def\csb{{\chi\rm{SB}}}

\def\hmu{\hat{\mu}}

\def\f0{\text{\boldmath$\varphi$}}
\def\h2{\mathfrak{h}}
\def\vol{{\rm vol}}

\catcode`\@=12

\begin{document}

%%%
%%%%%% text starts here
%%%%%%%%%

\title{\bf The quest for a  conifold conformal order}

\date{April 29, 2022}
%\date\today

\author{
Alex Buchel\\[0.4cm]
\it $ $Department of Physics and Astronomy\\ 
\it University of Western Ontario\\
\it London, Ontario N6A 5B7, Canada\\
\it $ $Perimeter Institute for Theoretical Physics\\
\it Waterloo, Ontario N2J 2W9, Canada
}

\Abstract{The holographic duality between cascading gauge theory and type IIB
supergravity on warped deformed conifold with fluxes reveals exotic thermal
phases with nonzero expectation values of certain operators, persistent
to high temperatures. These phases, in the limit of vanishing the strong
coupling scale of the cascading gauge theory, would realize thermal ordered
conformal phases in $\reals^{3,1}$ relativistic QFT.
We find that the dual Klebanov-Strassler/Klebanov-Tseytlin black
branes in this limit are outside the regime of the supergravity approximation,
rendering the construction of such conformal ordered states
unreliable. While we have been able to construct
conformal order in {\it phenomenologically deformed}
effective theory of type IIB supergravity
reduced on warped deformed conifold with fluxes, the removal of the
deformation parameter causes the destruction of the thermal conformal ordered
phases. Once again, we find that the holographic models with the
conformal ordered phases are in the String Theory swampland.
}

\makepapertitle

\body

\version\versionno
\tableofcontents

\section{Introduction}\label{intro}

{\it Conformal order} stands for exotic thermal phases of
conformal field theories (CFTs), characterized with 
nonzero one-point correlation function(s) of certain operator(s)
\cite{Chai:2020zgq,Buchel:2009ge,Buchel:2020thm,Buchel:2020jfs,Buchel:2020xdk,Chaudhuri:2020xxb,Chai:2021djc,Chaudhuri:2021dsq,Buchel:2021ead,Chai:2021tpt}.  
For a CFT$ _{d+1}$ in Minkowski space-time $\reals^{d,1}$ the existence
of the ordered phases implies that there are at least two distinct
thermal phases:
\begin{equation}
\frac{\calf}{T^{d+1}}=-\calc\ \times\
\begin{cases}
1,\qquad T^{-\Delta_i}\langle\calo_{\Delta_i}\rangle=0\,,\\
\kappa,\qquad T^{-\Delta_i}\langle\calo_{\Delta_i}\rangle=\gamma_i\ne 0\,,\ 
\end{cases}
\eqlabel{phd}
\end{equation}
where $\calf$ is the free energy density, $T$ is the temperature,
$\calc$ is a positive constant
proportional to the central charge of the theory, and $\{\calo_{\Delta_i}\}$ is the
set of the order parameters with the conformal dimension
spectrum $\{\Delta_i\}$. The parameters $\kappa$ and $\{\gamma_i\}$
characterizing the thermodynamics of 
the ordered phase are  necessarily constants.
Conformal order can realize spontaneous breaking of 
discrete \cite{Buchel:2009ge,Chai:2020zgq,Buchel:2020thm} of continuous 
\cite{Chai:2021tpt} global symmetries; but it does not have to be the case: in
the model we discussed in \cite{Buchel:2021ead}, the conformal order parameter is not
associated with spontaneous breaking of any global symmetry\footnote{This is strictly true in the context of the effective
Kaluza-Klein reduced holographic models. The nonvanishing
scalar expectation value would signal the spontaneous breaking of the
global symmetry of the compactification manifold
upon uplift to 10d supergravity. Unfortunately,
no model with a thermal conformal order has been constructed
in top-down holography yet.}.

From \eqref{phd}, note that
when $\kappa>1$ ($\kappa<1$), the symmetry broken phase dominates
(is subdominant) both
in the canonical and the microcanonical ensembles.
Irrespectively
of the value, provided $\kappa>0$, the symmetry broken phase is
thermodynamically stable. It is difficult to compute directly in
a CFT the values $\{\kappa, \gamma_i\}$, thus establishing 
the presence and the (in)stability of the ordered phase. 
Rather, the authors of
\cite{Chai:2020zgq,Chaudhuri:2020xxb,Chai:2021djc,Chaudhuri:2021dsq}
established the instability of the {\it disordered} 
thermal phases in discussed CFTs. The condensation of the
identified unstable mode then leads to $\langle\calo_{\Delta_i}\rangle\ne 0$
for the new equilibrium thermal state --- the conformal order.

Conformal order states are very interesting in the context of
holography \cite{Maldacena:1997re,Aharony:1999ti},
as they imply the  existence of the dual black branes in a Poincare patch of 
asymptotically $AdS_{d+2}$ bulk geometry that violate the no-hair theorem.
In \cite{Buchel:2021ead} we proved a theorem that the disordered
conformal thermal states are always stable in dual holographic models
of Einstein gravity with multiple scalars.
Thus, the mechanism for the conformal order presented
in \cite{Chai:2020zgq} is not viable in these holographic models.

The first holographic model of the conformal order was discovered (though not appreciated in this
context prior to the QFT construction \cite{Chai:2020zgq})
purely by accident in \cite{Buchel:2009ge}. The general framework for constructing
holographic conformal order, the  {\it phenomenologically deformed} effective
theory, was presented in \cite{Buchel:2020xdk} --- we now review those
arguments as they will be utilized in this paper\footnote{All the known constructions of the
holographic conformal order can be understood within this framework.}. 
Consider a top-down holographic model, dual to\footnote{Extensions to
AdS$ _{d+2}$/CFT$ _{d+1}$ models with multiple order parameters
$\calo_{\Delta_i}$ is trivial ---  such  models
will be studied in section \ref{ksdef}.}
a CFT$ _4$ with a single operator
$\calo_\Delta$ of a dimension $\Delta$. The five-dimensional
gravitational bulk effective action takes the form 
\begin{equation}
S_5=\frac{c}{2\pi^2 L^3} \int_{\calm_{5}} \vol_{\calm_5} \biggl\{R-\frac 12 (\nabla\phi)^2-\calp[\phi]
\biggr\}\,,
\eqlabel{ar1}
\end{equation}
where $c$ is the central charge of the boundary CFT, $\phi$ is the gravitational bulk scalar dual to
the order parameter $\calo_\Delta$, and the scalar potential $\calp[\phi]$ is  
\begin{equation}
\calp[\phi]=-\frac{12}{L^2}+\frac{\Delta(\Delta-4)}{2L^2} \phi^2+\calo(\phi^3)\,.
\eqlabel{ar2}
\end{equation}
The $\calo(\phi^0)$ term
in \eqref{ar2} is a negative cosmological constant, setting the radius of the
asymptotically AdS$ _5$ bulk geometry to $L$; the mass term,
\begin{equation}
L^2 m^2\equiv \Delta (\Delta-4)\,, 
\eqlabel{ar3}
\end{equation}
represents the standard encoding of the dimension of the order parameter
$\calo_\Delta$ \cite{Gubser:1998bc,Witten:1998qj}.
In what follows we set $L=1$.
It is vital that in real holographic models (contrary to the phenomenological toys), the full scalar
potential $\calp[\phi]$ is nonlinear. Unfortunately, holography is not understood at the level
where given a boundary CFT, with a spectrum of gauge invariant operators, we can engineer/compute
the scalar potential. In specific holographic examples, like the $\caln=2^*$ correspondence
\cite{Pilch:2000ue,Buchel:2000cn,Evans:2000ct}, the cascading gauge theory
duality \cite{Klebanov:2000hb,Herzog:2001xk} or the Maldacena-Nunez model \cite{Maldacena:2000yy},
the scalar potential is computed from the realization of the duality correspondence in
type IIB supergravity. The construction of the holographic conformal order proposed in
\cite{Buchel:2020xdk} relies on (and is applicable to) models where the leading nonlinear
correction is unbounded from below\footnote{This feature is ubiquitous in top-down holographic examples.} 
along certain directions on the scalar manifold. To be specific, we assume
that
\begin{equation}
\calp[\phi]=-{12}+\frac{\Delta(\Delta-4)}{2} \phi^2-\frac{s^2}{2}\ \phi^n +\calo(\phi^{n+1})\,,
\eqlabel{ar4}
\end{equation}
for some constant $s$ and an integer $n\ge 3$. Note that $\calo(\phi^n)$-truncated 
scalar potential is unbounded as $\phi\to +\infty$. Of course, the full scalar potential
$\calp[\phi]$ might be bounded, but  this is not important for the
perturbative construction of the thermal conformal order in the {\it phenomenologically deformed} model.
The phenomenologically deformed model is defined as a holographic correspondence
where the top-down scalar potential $\calp[\phi]$ is deformed as 
\begin{equation}
\calp[\phi]\ \longrightarrow\ \calp^b[\phi]\equiv -{12}+\frac{\Delta(\Delta-4)}{2} \phi^2
+b \biggl(\calp[\phi]+12-\frac{\Delta(\Delta-4)}{2} \phi^2\biggr)\,,
\eqlabel{ar5}
\end{equation}
with the constant deformation parameter $b$ being positive. The claim of \cite{Buchel:2020xdk},
see also appendix \ref{pertorder}, is that the thermal conformal order always exists
in the limit $b\to +\infty$, when the thermal ordered phase is holographically
realized as AdS-Schwarzschild black brane, with a perturbatively small ``scalar hair''
\begin{equation}
\phi\ \propto\ \left(\frac 1b\right)^{\frac{1}{n-2}}\qquad \Longleftrightarrow\qquad T^{-\Delta}\langle\calo_\Delta\rangle=\gamma \propto \left(\frac 1b\right)^{\frac{1}{n-2}} \,.
\eqlabel{ar6}
\end{equation}
The existence of the thermal conformal order in real holography then boils to the question 
whether this perturbative constructions survives as $b$ decreases from $+\infty$ to $1$, since
\begin{equation}
\lim_{b\to 1_+} \calp^b[\phi]\ =\ \calp[\phi] \,.
\eqlabel{ar7}
\end{equation}

In this paper we continue the quest for constructing the thermal conformal order in String Theory
holography. Our focus is on  top-down holographic dualities between regular and fractional
D3-branes on a conifold, the simplest non-compact Calabi-Yau threefold \cite{Candelas:1989js},
and $\caln=1$ supersymmetric gauge theories - the Klebanov-Witten (KW) \cite{Klebanov:1998hh}
and the Klebanov-Strassler (KS) \cite{Klebanov:2000hb} models. There are two reasons for this
choice:
\begin{itemize}
\item Analysis of the phenomenological model \cite{Buchel:2009ge} revealed for the
first time the exotic thermal phases in a holographic system, which are associated with the
spontaneous breaking of a discrete symmetry, and persist to arbitrary high temperature.
As $T\to \infty$, the fact that the model of  \cite{Buchel:2009ge} was non-conformal becomes
irrelevant since for any fixed mass scale $m/T\to 0$. In this way one can obtain a
holographic thermal conformal order, as emphasized in \cite{Buchel:2020thm}. Holographic duality
between type IIB supergravity on warped deformed conifold with fluxes and the KS cascading
gauge theory also reveals the exotic thermal phase \cite{Buchel:2018bzp} --- the deconfined phase with spontaneously broken
$U(1)_R$ chiral symmetry, that exists only above certain critical temperature
$T_\csb$. Like the model in \cite{Buchel:2009ge}, the cascading gauge theory is non-conformal,
and has a strong coupling scale $\Lambda$. It is natural to explore this exotic phase in
the limit $\Lambda/T\to 0$, and potentially obtain a top-down holographic model
of the thermal conformal order. We discuss this in section \ref{ksktbhs}.
\item The limit of $\Lambda/T\to 0$ in the above example effectively removes
the fractional D3-branes from the holographic model. On the gravity side one ends with 
type IIB supergravity on warped deformed conifold with the self-dual five-form flux.
The corresponding boundary gauge theory is $\caln=1$ superconformal KW model.
The gravitational bulk scalars  encode the gauge invariant operators
(potential  conformal order parameters). As we review in section \ref{ksdef},
the resulting holographic model (when the conifold is not deformed)
is a universal example of AdS$ _5/$CFT$ _4$ duality on warped and squashed
Sasaki-Einstein manifolds \cite{Cassani:2010uw}. Thus, analysis
of the conformal order on the conifold will cover all such cases, see sections
\ref{modeli} and \ref{modelii}.
In the case of the deformed conifold, we will have potentially the first example
of the holographic conformal order with spontaneously broken continuous symmetry,
see section \ref{modeliii}.
\end{itemize}

We summarized our results in section \ref{conclude}.

\section{Exotic phases of the cascading gauge theory at high temperature}\label{ksktbhs}

Thermodynamics of the Klebanov-Strassler cascading gauge theory \cite{Klebanov:2000hb}
has by now a long history \cite{Buchel:2000ch,Buchel:2001gw,Gubser:2001ri,
Aharony:2005zr,Aharony:2007vg,Buchel:2009bh,Buchel:2010wp,Buchel:2018bzp,Buchel:2021yay}.
We refer the reader to a recent comprehensive review \cite{Buchel:2021yay},
and focus here on the results only.

Cascading gauge theory is $\caln=1$ supersymmetric $SU(N+M)\times SU(N)$ gauge theory with
pairs of chiral multiplets $A_k$ and $B_\ell$, $k,\ell=1,2,$ in the bifundamental
$(N+M,\overline{N})$ and $(\overline{N+M},{N})$ representations, and the
superpotential
\begin{equation}
\calw\ \propto\ \epsilon^{ij}\epsilon^{k\ell}\ {\rm Tr}\left(A_iB_kA_jB_\ell\right)\,.
\eqlabel{wks}
\end{equation}
The theory is not conformal, and the gauge couplings 
$g_1$ and $g_2$, of the gauge group factors $SU(N+M)$
and $SU(N)$ correspondingly, run with the renormalization group scale $\hat{\mu}$,
\begin{equation}
\begin{split}
&\frac{d}{d\ln(\hmu/\Lambda)}\ \frac{8\pi^2}{g_1^2}=3(N+M)-2 N(1-\gamma)\,,\\
&\frac{d}{d\ln(\hmu/\Lambda)}\ \frac{8\pi^2}{g_2^2}=3N-2 (N+M)(1-\gamma)\,,
\end{split}
\eqlabel{rgrunning}
\end{equation}
where $\gamma$ is the anomalous dimension of operators $\tr A_iB_j$ and $\Lambda$ is the strong coupling scale of
the cascading gauge theory.
To leading order in $M/N$, $\gamma=-\frac12$ \cite{Klebanov:1998hh,Klebanov:2000hb}, so that
\begin{equation}
 \frac{8\pi^2}{g_1^2}- \frac{8\pi^2}{g_2^2}= 6 M \ln \frac{\hmu}{\Lambda}\ \times\ \biggl(1+\calo({M}/{N})\biggr)\,,
\eqlabel{rg1}
\end{equation}
while the sum of the gauge couplings is constant along the RG flow
\begin{equation}
 \frac{8\pi^2}{g_1^2}+ \frac{8\pi^2}{g_2^2}={\rm const}\,.
\eqlabel{rg2}
\end{equation}

\begin{figure}[t]
\begin{center}
\psfrag{t}[cc][][1][0]{${T}/{\Lambda}$}
\psfrag{f}[bb][][1][0]{$\hat{\calf}$}
\psfrag{k}[tt][][1][0]{$\ln \frac{\calk_\csb}{\calk}$}
\includegraphics[width=3in]{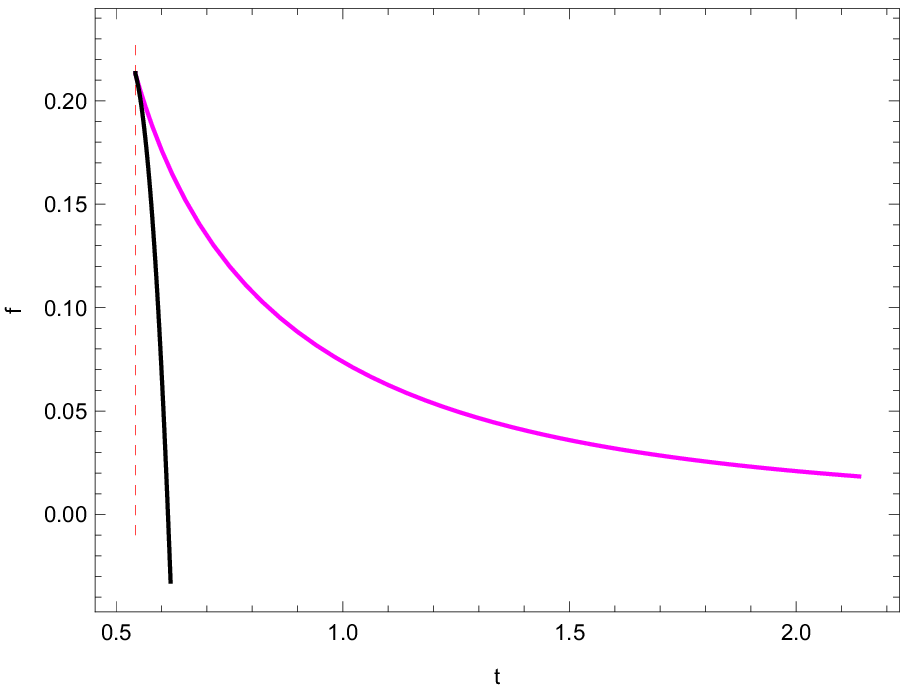}
\includegraphics[width=3in]{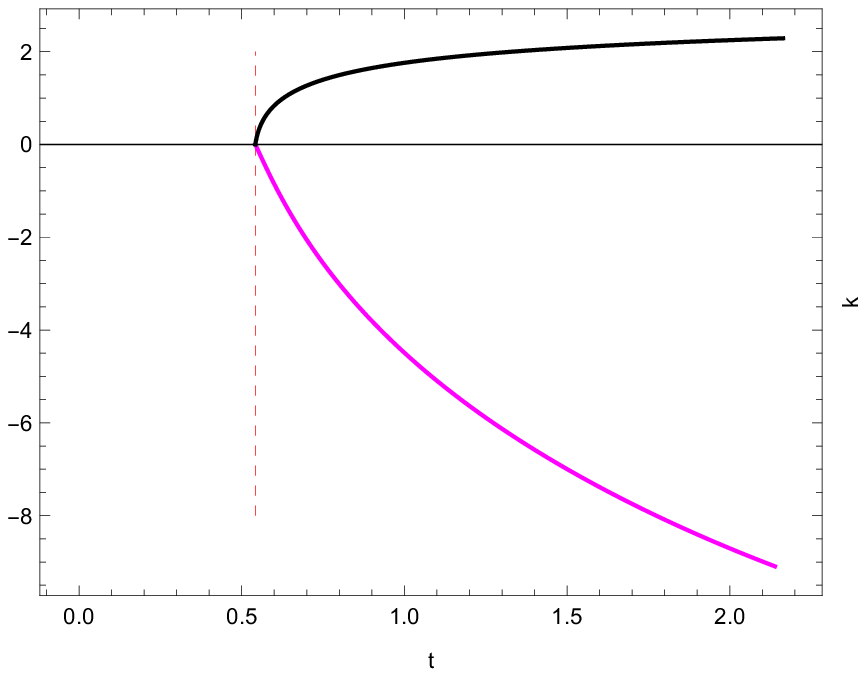}
\end{center}
  \caption{The left panel: the reduced free energy density
  $\hat\calf$ \eqref{fhatdef} of the
  deconfined chirally symmetric phase (the black curve), and the 
deconfined (ordered) phase with spontaneously broken chiral
symmetry (the magenta curve)
of the cascading gauge theory plasma. The vertical dashed red lines
indicate $T_\csb$ \eqref{tcsb}. The ordered phase is exotic:
it extends for $T>T_\csb$. The right panel: the Kretschmann scalar $\calk$
for the corresponding phases computed for the dual black branes
at the horizon. We use $\calk_\csb=\calk(T=T_\csb)$.} \label{kskts}
\end{figure}

\begin{figure}[t]
\begin{center}
\psfrag{t}[cc][][1][0]{${T}/{\Lambda}$}
\psfrag{f}[bb][][1][0]{$\hat{\calf}$}
\psfrag{k}[tt][][1][0]{$\ln \frac{\calk_\csb}{\calk}$}
\includegraphics[width=3in]{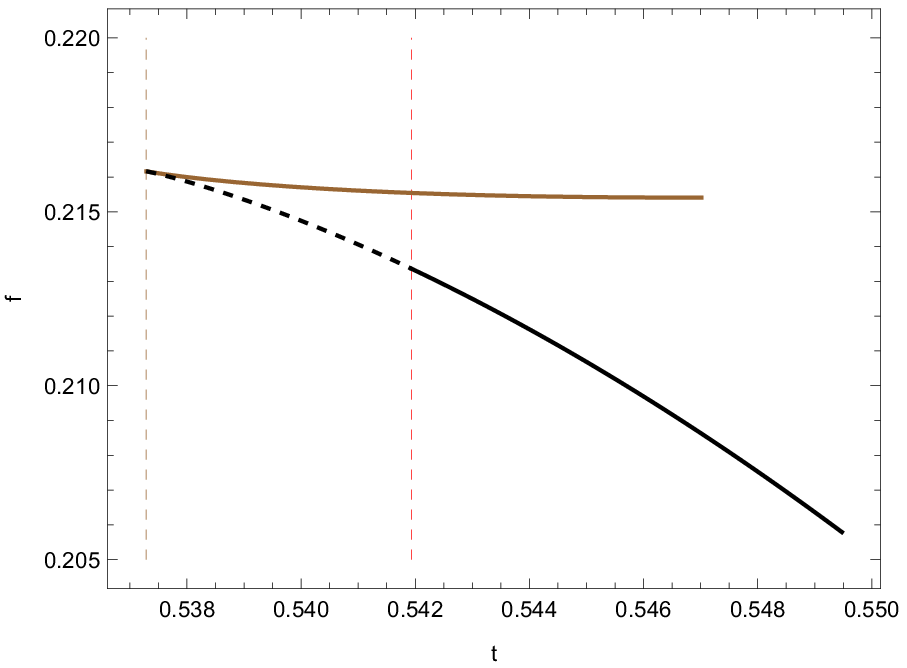}
\includegraphics[width=3in]{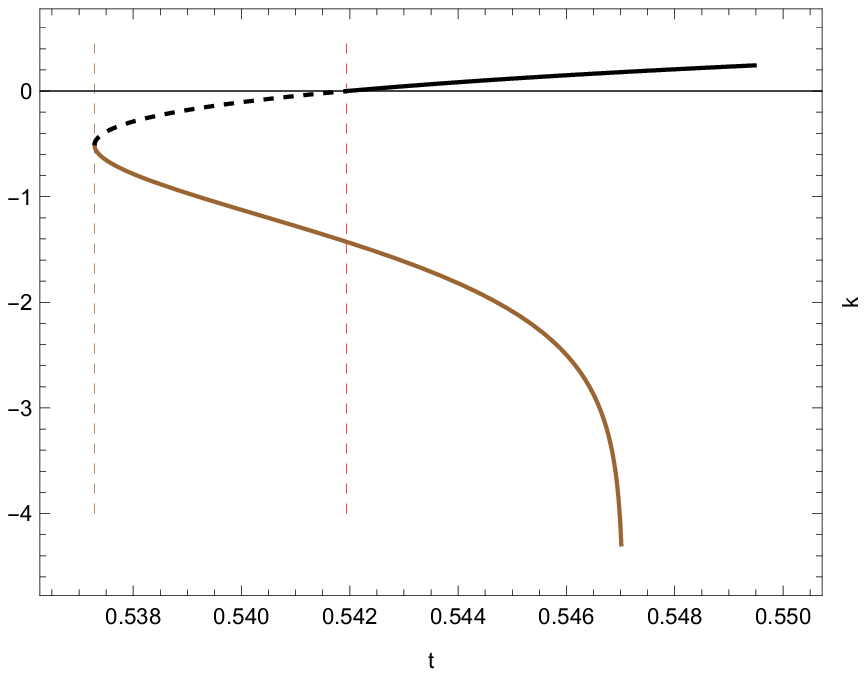}
\end{center}
  \caption{The left panel: the reduced free energy density
  $\hat\calf$ \eqref{fhatdef} of the
  deconfined chirally symmetric phase with the positive specific heat (the black curves), and the 
deconfined chirally symmetric phase with the negative specific heat (the brown curve)
of the cascading gauge theory plasma. The vertical dashed brown lines
indicate $T_u$ \eqref{tudef}. The negative specific heat phase is exotic:
it extends for $T>T_u$ with ever increasing thermal expectation values of
certain gauge invariant operators. The right panel: the Kretschmann scalar $\calk$
for the corresponding phases computed for the dual black branes
at the horizon. As in fig.~\ref{kskts}, we normalize the Kretschmann scalar to its
value at $T_\csb$.
} \label{ktukts}
\end{figure}

The thermal phase diagram of the theory is rich:
\nxt At large temperatures, $T\gg \Lambda$, its thermal equation of state is that of a conformal theory
with the effective temperature-dependent central charge \cite{Buchel:2000ch}, {\it e.g},
the free energy density $\calf$ takes the form
\begin{equation}
\calf\ \propto\ -c_{eff}(T)\ T^4\,,\qquad c_{eff}\propto M^4 \ln^2 \frac T\Lambda\,.
\eqlabel{flarget}
\end{equation}
\nxt At
\begin{equation}
T=T_{c}=0.614(1)\Lambda\,,
\end{equation}
the theory undergoes the first-order confinement/deconfinement phase transition \cite{Aharony:2007vg}; precisely at the
transition point the free energy density vanishes,
\begin{equation}
\hat{\calf}\equiv \frac{2^6\pi^4}{3^5M^4}\ \frac{\calf}{\Lambda^4}\bigg|_{T=T_c}=0\,,
\eqlabel{fhatdef}
\end{equation}
where the first equality introduces dimensionless quantity $\hat\calf$ we use to describe cascading
gauge theory thermodynamics in the canonical ensemble.
\nxt The next critical temperature is \cite{Buchel:2010wp}
\begin{equation}
T_\csb=0.541(9)\Lambda\,,
\eqlabel{tcsb}
\end{equation}
represented by the vertical dashed red lines in figs.~\ref{kskts} and \ref{ktukts}.
For $T>T_{\csb}$ the deconfined phase, represented by the
solid black curves in figs.~\ref{kskts} and \ref{ktukts},
is perturbatively stable to $U(1)_R\to \zet_2$ chiral symmetry breaking fluctuations.  The deconfined states of the cascading gauge theory plasma
unstable to chiral symmetry breaking fluctuations
are represented by the dashed solid black curves in figs.~\ref{kskts} and \ref{ktukts}.
\nxt Similar to the model of \cite{Buchel:2009ge}, the deconfined phase with spontaneously broken chiral
symmetry, the  solid magenta curves in fig.~\ref{kskts},  is {\color{magenta}{\it exotic}}: it exists for $T>T_\csb$ and is
realized holographically as the Klebanov-Strassler black brane \cite{Buchel:2018bzp}.
If we could extend this phase for $\frac{T}{\Lambda}\to \infty$, by analogy with \cite{Buchel:2020thm},
we would have realized the conformal order on the conifold. Alas, our numerics allowed the construction
of this exotic phase in the range 
\begin{equation}
\frac{T}{\Lambda}=\frac{T_\csb}{\Lambda}\ \times \biggl\{1\cdots 3.9466\biggr\} \,.
\eqlabel{magentarange}
\end{equation}
There is a practical and a conceptual reason for this:
\begin{itemize}
\item from the practical perspective, certain normalizable
components of the scalar fields near the boundary become too large for a
reliable numerics\footnote{In the notations of \cite{Buchel:2021yay}, {\it e.g.}, it is
the coefficient $f_{c,8,0}$ --- see (A.55) there.};
\item the conceptual reason
causing the above growth is the following: as the temperature increases, the curvature of the dual Klebanov-Strassler black brane evaluated
at the horizon grows, and the construction becomes unreliable in the supergravity approximation. Specifically, in the right panel of
fig.~\ref{kskts} we present the value of the Kretschmann scalar
of the KS black brane horizon, for the exotic phase (the magenta curve)
and the deconfined chirally symmetric phase (the black curve), relative to the
value of the Kretschmann scalar $\calk_{\csb}$ evaluated at $T=T_\csb$.
Over the range \eqref{magentarange}, the Kretschmann scalar corresponding
to the exotic phase changes as
\begin{equation}
\calk=\calk_{\csb}\times \biggl\{1\cdots 8890\biggr\}\,.
\eqlabel{kmagenta}
\end{equation}
Thus, we conclude that there can not be a reliable conformal order
on the warped deformed conifold with fluxes, arising from the
$T\to\infty$ limit of the Klebanov-Strassler black branes --- such black branes
are outside the validity of the supergravity approximation.
\item Lastly, there is a terminal temperature \cite{Buchel:2009bh}
\begin{equation}
T_u=0.537(3)\Lambda \,,
\eqlabel{tudef}
\end{equation}
represented by the vertical dashed brown lines in fig.~\ref{ktukts},
for the deconfined chirally symmetric states of the cascading gauge theory
plasma --- these states exist only for $T\ge T_u$. In the vicinity of
$T_u$, there are two branches of states: the stable branch with respect to
the energy density fluctuations (the black curves in fig.~\ref{ktukts}),
and the unstable
branch (the brown curves in fig.~\ref{ktukts}). On the former branch the specific heat is positive,
while it is negative on the latter branch,
explaining the (in)stability to the energy density fluctuations
\cite{Buchel:2005nt}. The unstable
branch of the deconfined chirally symmetric states is {\color{brown} exotic}:
it extends for $T>T_u$ with the ever increasing values of the thermal expectation
values of $\calo_{\Delta=\{4,6,8\}}$ , however, much like in the case of the exotic phase with
the chiral symmetry breaking, we fail to extend it  as $T\to \infty$. We constructed
the latter states for fairly narrow temperature range:
\begin{equation}
\frac{T}{\Lambda}=\frac{T_u}{\Lambda}\times\biggl\{1\cdots 1.0181\biggr\}\,.
\eqlabel{brownrange}
\end{equation}
As the temperature of the exotic deconfined chirally symmetric phase increases,
the curvature of the corresponding dual Klebanov-Tseytlin black brane evaluated at the
horizon rapidly grows,
see the right panel of fig.~\ref{ktukts}. Over the range \eqref{brownrange},
the Kretschmann scalar corresponding
to this exotic phase (the brown curve) changes as
\begin{equation}
\calk=\calk_{\csb}\times \biggl\{1.669\cdots 73.1\biggr\}\,.
\eqlabel{kbrown}
\end{equation}
On the contrary, the deconfined chirally symmetric phase with the positive specific heat
(the black curves in fig.~\ref{ktukts}) can be extended as $\frac T\Lambda\to \infty$ --- however it is
not exotic, as the thermal expectation
values of $\calo_{\Delta=\{4,6,8\}}$ operators vanish \cite{Aharony:2007vg} in this limit, and
we end up with the log-dressed conformal equation of state \eqref{flarget}.
Thus, we conclude that there can not be a reliable conformal order
on the warped squashed conifold with fluxes, arising from the
$T\to\infty$ limit of the Klebanov-Tseytlin black branes with the negative specific heat ---
such black branes are outside the validity of the supergravity approximation.
\end{itemize}

\section{Effective theories and the conformal order on the conifold}\label{ksdef}

Consistent truncation in the $SU(2)\times SU(2)\times \zet_2$ invariant sector of type IIB supergravity on
warped deformed conifold with fluxes  to a five dimensional manifold $\calm_5$  was derived
in \cite{Buchel:2010wp}\footnote{See \cite{Buchel:2021yay} for a recent comprehensive review.}:
\begin{equation}
\begin{split}
S_5\biggl[\hg_{\mu\nu},&\Omega_{i=1\cdots 3},\Phi,h_{i=1\cdots 3}\,,\,\{P,\Omega_0\}\biggr]= \frac{108}{16\pi G_5} 
\int_{\calm_5} \hat{{\rm vol}}_{\calm_5}\ \Omega_1 \Omega_2^2\om_3^2\\
&\times \biggl\lbrace 
 R_{10}-\frac 12 \left(\hat{\nabla} \Phi\right)^2
-\frac 12 e^{-\Phi}\left(\frac{(h_1-h_3)^2}{2\om_1^2\om_2^2\om_3^2}+\frac{1}{\om_3^4}\left(\hat{\nabla} h_1\right)^2
+\frac{1}{\om_2^4}\left(\hat{\nabla} h_3\right)^2\right)
\\
&-\frac 12 e^{\Phi}\left(\frac{2}{\om_2^2\om_3^2}\left(\hat{\nabla} h_2\right)^2
+\frac{1}{\om_1^2\om_2^4}\left(h_2-\frac P9\right)^2
+\frac{1}{\om_1^2\om_3^4} h_2^2\right)
\\
&-\frac {1}{2\Omega_1^2\Omega_2^4\om_3^4}\left(4\Omega_0+ h_2\left(h_3-h_1\right)+\frac 19 P h_1\right)^2
\biggr\rbrace.
\end{split}
\eqlabel{5action}
\end{equation}
It is a functional of the a five-dimensional metric $\hg_{\mu\nu}$ on $\calm_5$,
\begin{equation}
ds_{5}^2 =\hg_{\mu\nu}(y) dy^{\mu}dy^{\nu}\,,
\eqlabel{5met}
\end{equation}
scalars $\Omega_{i=1\cdots 3}$ describing the warping and the deformation of the conifold,
a dilaton $\Phi$, scalars $h_{i=1\cdots 3}$ parameterizing the 3-form fluxes, a constant
parameter $\Omega_0$ (necessary to define the self-dual 5-form flux),
and a topological parameter $P$,
\begin{equation}
\frac{2P}{9\alpha'}\equiv M\ \in\ \zet\,,
\eqlabel{defp}
\end{equation}
related to the number of fractional D3 branes on the conifold. Finally,
$R_{10}$ is the Ricci scalar of the
ten-dimensional type IIB metric, obtained from uplifting \eqref{5met},
\begin{equation}
\begin{split}
R_{10}=\hat{R}_5&+\left(\frac{1}{2\om_1^2}+\frac{2}{\om_2^2}+\frac{2}{\om_3^2}-\frac{\om_2^2}{4\om_1^2\om_3^2}
-\frac{\om_3^2}{4\om_1^2\om_2^2}-\frac{\om_1^2}{\om_2^2\om_3^2}\right)-2\hat{\Box} \ln\left(\om_1\om_2^2\om_3^2\right)\\
&-\biggl\{\left(\hat{\nabla}\ln\om_1\right)^2+2\left(\hat{\nabla}\ln\om_2\right)^2
+2\left(\hat{\nabla}\ln\om_3\right)^2+\left(\hat{\nabla}\ln\left(\om_1\om_2^2\om_3^2\right)\right)^2\biggr\}\,,
\end{split}
\eqlabel{ric5}
\end{equation}
and $\hat{R}_5$ is the five-dimensional Ricci scalar of the metric \eqref{5met}.   

We find it convenient to rewrite the action \eqref{5action} in five-dimensional Einstein frame.
The latter is achieved with the following rescaling
\begin{equation}
\hg_{\mu\nu}\ \to\ \Om^2 g_{\mu\nu}\,,\qquad \Om^{-3}\equiv 108\ \Om_1\Om_2^2\Om_3^2\,,
\eqlabel{5drescale}
\end{equation}
leading to
\begin{equation}
\begin{split}
108\ \sqrt{-\hg}\ \Om_1\Om_2^2\Om_3^2\ \hat{R}_5\ =\ \sqrt{-g}\biggl(\
R-8\Box \ln\Om-12\left(\nabla\ln\Om\right)^2
\ \biggr)\,.
\end{split}
\eqlabel{EH}
\end{equation}
Further introducing
\begin{equation}
\Om_1=\frac13\ e^{f-4w}\,,\qquad \Om_2=\frac{1}{\sqrt{6}}\ e^{f+w+\lambda}\,,\qquad \Om_3=\frac{1}{\sqrt{6}}\ e^{f+w-\lambda}\,,
\eqlabel{omscalars}
\end{equation}
the five-dimensional effective action becomes
\begin{equation}
\begin{split}
&S_5=\frac{1}{16\pi G_5}\int_{\calm_5} {\rm vol}_{\calm_5}\ \biggl\{
R-\frac{40}{3} \left(\nabla f\right)^2-20 \left(\nabla w\right)^2-4 \left(\nabla \lambda\right)^2
-\frac 12\left(\nabla \Phi\right)^2\\
&-18 e^{-4 f -4w-\Phi}\biggl[e^{4\lambda} \left(\nabla h_1\right)^2+e^{-4\lambda} \left(\nabla h_3\right)^2\biggr]-36 e^{-4f -4w+\Phi} \left(\nabla h_2\right)^2-\calp_{flux}-\calp_{scalar}
\biggr\}\,,
\end{split}
\eqlabel{eaeh}
\end{equation}
where
\begin{equation}
\begin{split}
\calp_{flux}=&81 e^{-\frac{28}{3}f +4w-\Phi}(h_1-h_3)^2+162  e^{-\frac{28}{3}f +4w+\Phi}
\biggl[e^{-4\lambda}\left(h_2-\frac 19 P\right)^2+e^{4\lambda} h_2^2\biggr]\\
&+72 e^{-\frac{40}{3}f}\biggl[h_1(P-9 h_2)+9 h_2 h_3 +36\Om_0\biggr]^2\,,
\end{split}
\eqlabel{pflux}
\end{equation}
\begin{equation}
\calp_{scalar}=4 e^{-\frac{16}{3}f-12 w}-24e^{-\frac{16}{3}f-2w}\cosh(2\lambda)-\frac 92 e^{-\frac{16}{3}f+8w}
\biggl(1-\cosh(4\lambda)\biggr)\,.
\eqlabel{pscalar}
\end{equation}
Consistent truncation of \eqref{eaeh}, \ie
\begin{equation}
\lambda=0\,,\qquad h_1=h_3=\frac 1P\left(\frac{K}{12}-36\Om_0\right)\,,\qquad h_2=\frac{P}{18}\,,
\eqlabel{u1sym}
\end{equation}
produces effective action of $SU(2)\times SU(2)\times U(1)$ invariant sector of type IIB
supergravity on warped squashed conifold with fluxes derived in \cite{Buchel:2005cv}. 

The boundary holographic dual represented by \eqref{eaeh} is the $\caln=1$ supersymmetric
$SU(N+M)\times SU(N)$ cascading gauge theory, often referred to as a Klebanov-Strassler
gauge theory \cite{Klebanov:2000hb}. This theory is not conformal, and has a strong coupling
scale $\Lambda$,
\begin{equation}
\Lambda^2=\frac{\sqrt 2}{P^2 g_s}\ e^{-\frac{K_0}{P^2 g_s}}\,,
\eqlabel{dellambda}
\end{equation}
where $g_s$ is the asymptotic value of the string coupling
constant\footnote{It can always be fixed to $g_s=1$.}, and
$K_0$ is a parameter that can always be adjusted
(using the scaling symmetries of the holographic radial coordinate)
to a fixed positive value. The precise definition of $K_0$ can be found in
\cite{Bena:2019sxm,Buchel:2021yay}. We are interested here in the conformal
holographic model, thus we must take the limit $\Lambda\to 0$, which is equivalent 
to sending $P\to 0$. Finding the conformal order in the model with 7 scalars is
a daunting task --- so\footnote{While it is conceivable that
other general constructions of the conformal order might exist in the model,
finding them without any guidances is a lost cause.}
we will follow the framework developed in \cite{Buchel:2020xdk},
and reviewed in appendix \ref{pertorder}.
\nxt As a first step, we truncate the general potential $\calp_{flux}+\calp_{scalar}$ in
\eqref{eaeh} to scalars that enter nonlinearly, and have ``unbounded'' directions on the
field space manifold, at least close to the origin. Notice that $h_i$, encoding
the 3-form fluxes on the conifold, enter quadratically and generate always nonnegative
contribution to the scalar potential, \ie $\calp_{flux}\ge 0$. Thus, in addition to
$P=0$, we set $h_i\equiv 0$. Then,
\begin{equation}
\calp_{flux}=72 (36 \Om_0)^2\ e^{-\frac{40}{3}f} =8 e^{-\frac{40}{3}f}\,,
\eqlabel{pfluxc}
\end{equation}
where in the second equality we set a constant parameter $\Om_0=\frac{1}{108}$ to
ensure that the radius of the asymptotically AdS$ _5$ geometry is $L=1$. 
\nxt In the absence of 3-form fluxes the dilaton becomes an exact modulus, and we set
\begin{equation}
\Phi\equiv 0\,.
\eqlabel{setdil}
\end{equation}
\nxt We now arrive at the effective action we use to analyze conformal order on the warped
deformed conifold:
\begin{equation}
\begin{split}
&S_5=\frac{c_{KW}}{2\pi^2} \int_{\calm_5} {\rm vol}_{\calm_5}\ \biggl\{
R-\frac{40}{3} \left(\nabla f\right)^2-20 \left(\nabla w\right)^2-4 \left(\nabla \lambda\right)^2
-\calp[f,w,\lambda]
\biggr\}\,,
\end{split}
\eqlabel{eaco}
\end{equation}
where
\begin{equation}
c_{KW}=\frac{27N^2}{64}\,,
\eqlabel{ckw}
\end{equation}
is the central charge of $\caln=1$ superconformal $SU(N)\times SU(N)$ Klebanov-Witten  model  
\cite{Klebanov:1998hh}, and
\begin{equation}
\calp=4 e^{-\frac{16}{3}f-12 w}-24e^{-\frac{16}{3}f-2w}\cosh(2\lambda)-\frac 92 e^{-\frac{16}{3}f+8w}
\biggl(1-\cosh(4\lambda)\biggr)+8 e^{-\frac{40}{3}f}\,.
\eqlabel{pco}
\end{equation}
From \eqref{eaco} we obtain the following equations of motion
\begin{equation}
0=\Box f-\frac{3}{80}\ \frac{\del \calp}{\del f}\,,
\eqlabel{eom1}
\end{equation}
\begin{equation}
0=\Box w-\frac{1}{40}\ \frac{\del \calp}{\del w}\,,
\eqlabel{eom2}
\end{equation}
\begin{equation}
0=\Box \lambda-\frac{1}{8}\ \frac{\del \calp}{\del \lambda}\,,
\eqlabel{eom3}
\end{equation}
\begin{equation}
R_{\mu\nu}=\frac{40}{3}\ \del_\mu f\del_\nu f+20\ \del_\mu w\del_\nu w+4\ \del_\mu \lambda\del_\nu \lambda
+\frac 13 g_{\mu\nu}\ \calp\,.
\eqlabel{eom4}
\end{equation}

In the rest of this section we consider three conformal models, obtained by consistent
truncations of the effective action $S_5=S_5[f,w,\lambda]$ \eqref{eaco}:
\begin{itemize}
\item Model I:
\begin{equation}
S_5^I [f]\ \equiv\ S_5\bigg|_{w=\lambda\equiv0} \,;
\eqlabel{defm1}
\end{equation}
\item Model II:
\begin{equation}
S_5^{II} [f,w]\ \equiv\ S_5\bigg|_{\lambda\equiv0} \,;
\eqlabel{defm2}
\end{equation}
\item Model III:
\begin{equation}
S_5^{III} [f,w,\lambda]\ \equiv\ S_5 \,.
\eqlabel{defm3}
\end{equation}
\end{itemize}

\subsection{Conformal order in Model I}\label{modeli}

Conformal Model I realizes holographic dual to the KW gauge theory with a single
dimension $\Delta=8$ order parameter $\calo_8$, represented by the bulk scalar $f$.
Since this scalar represents the warping of the $T^{1,1}$ base of the singular conifold
\cite{Candelas:1989js}, the model is universal in the sense that any holographic duality
between a CFT$ _4$ and a type IIB supergravity on AdS$ _5\times$$Y_{ _5}$, where
$Y_5$ is a five-dimensional Sasaki-Einstein manifold, can be truncated to our Model I
\cite{Cassani:2010uw}.
In the latter case, the scalar $f$ represents the breathing mode of $Y_5$.

We follow appendix \ref{pertorder} to construct conformal order in Model I.
From \eqref{eaco} we find
\begin{equation}
\calp_I[f]= \calp[f,0,0]=-12+\frac{1280}{3} f^2-\frac{71680}{27}f^3+\calo(f^4)\,,
\eqlabel{p1}
\end{equation}
\ie the leading nonlinear term is $n=3$ (see \eqref{ar4}) and the unbounded direction is
along $f\to +\infty$. 
The $b$-deformed scalar potential takes the form, compare with \eqref{ar5},
\begin{equation}
\calp_I^b[f]=-12+\frac{1280}{3} f^2+b\biggl(-20e^{-\frac{16}{3}f}+8 e^{-\frac{40}{3}f}+12-\frac{1280}{3} f^2\biggr)\,.
\eqlabel{p1b}
\end{equation}
Using the radial coordinate as in \eqref{apx}, and introducing
\begin{equation}
c_2(x)=\frac{g(x)}{(2 x-x^2)^{1/4}}\,,
\eqlabel{defc2}
\end{equation}
we obtain the following equations of motion ($ '\equiv \frac{d}{dx}$)
\begin{equation}
\begin{split}
&0=f''+\frac{f'}{x-1}+ \biggl(
-\frac12 (f')^2+\frac{9h^2}{20}-\frac{9 (x^2-2 x+2)h}{40 x (1-x) (2-x)} +\frac{9}{80 x^2 (2-x)^2}
\biggr)\frac{\frac{\del\calp_I^b}{\del f}}{\calp_I^b}\,,
\end{split}
\eqlabel{eq1mod1}
\end{equation}
\begin{equation}
\begin{split}
&0=h'-4 h^2+\frac{(3 x^2-6 x+4) h}{x (1-x) (2-x)}+\frac{40}{9} (f')^2\,,
\end{split}
\eqlabel{eq2mod1}
\end{equation}
\begin{equation}
\begin{split}
&0=g'-h\ g\,.
\end{split}
\eqlabel{eq3mod1}
\end{equation}
Eqs.~\eqref{eq1mod1}-\eqref{eq2mod1} are solved subject to the boundary conditions:
\nxt as $x\to 0_+$, \ie at the AdS$ _5$ boundary
\begin{equation}
f= f_{2} x^2 +\calo(x^3)\,,\qquad h=-\frac{32}{9}f_2^2 x^3+\calo(x^4)\,,\qquad g=A
\left(1-\frac 89 f_2^2 x^4+\calo(x^5)\right)\,,
\eqlabel{mod1bc1}
\end{equation}
\nxt as $y\equiv (1-x)\to 0_+$, \ie at the horizon
\begin{equation}
f=f_0^h+\calo(y^2)\,,\qquad h=h_1^h y +\calo(y^3)\,,\qquad g=A\biggl(g_0^h+\calo(y^2)\biggr)\,.
\eqlabel{mod1bc2}
\end{equation}
Note that in total we have four parameters $\{f_2,f_0^h,h_1^h,g_0^h\}$,
along with an arbitrary constant $A$. They determine the thermal conformal order
of Model I, specifically,
\begin{equation}
\begin{split}
&s=\frac{2 c_{KW}}{\pi}\ A^3\left(g_0^h\right)^3\,,\qquad \calf=-\frac{sT}{4}\,,\qquad
T^{-8}\langle\calo_8\rangle=f_2\,,\\
&T^2=\frac{A^2\left(g_0^h\right)^2}{9\pi^2(1-2 h_1^h)}\biggl(
\left(320 (f^h_0)^2-9\right) (b-1)+3b\left(5e^{-\frac{16}{3}f^h_0}-2e^{-\frac{40}{3}f^h_0}\right)
\biggr)\,,
\end{split}
\eqlabel{data11}
\end{equation}
for the entropy density $s$, the free energy density $\calf$,  the temperature $T$,
and the thermal order parameter $\calo_8$.
Thus,  see \eqref{phd},
\begin{equation}
\kappa_I=27\left[\ \frac{\left(320 (f^h_0)^2-9\right) (b-1)+3b\left(5e^{-\frac{16}{3}f^h_0}-2e^{-\frac{40}{3}f^h_0}\right)}{1-2 h_1^1}\ \right]^{-3/2}\,,\ \  \gamma_{I,8}=f_2\,,
\eqlabel{kg1}
\end{equation}
where the subscript $ _I$ refers to Model I.

\begin{figure}[t]
\begin{center}
\psfrag{g}[cc][][1][0]{$1/\gamma_{I,8}$}
\psfrag{b}[bb][][1][0]{$b$}
\psfrag{i}[tt][][1][0]{$1/\gamma_{I,8}$}
\includegraphics[width=3in]{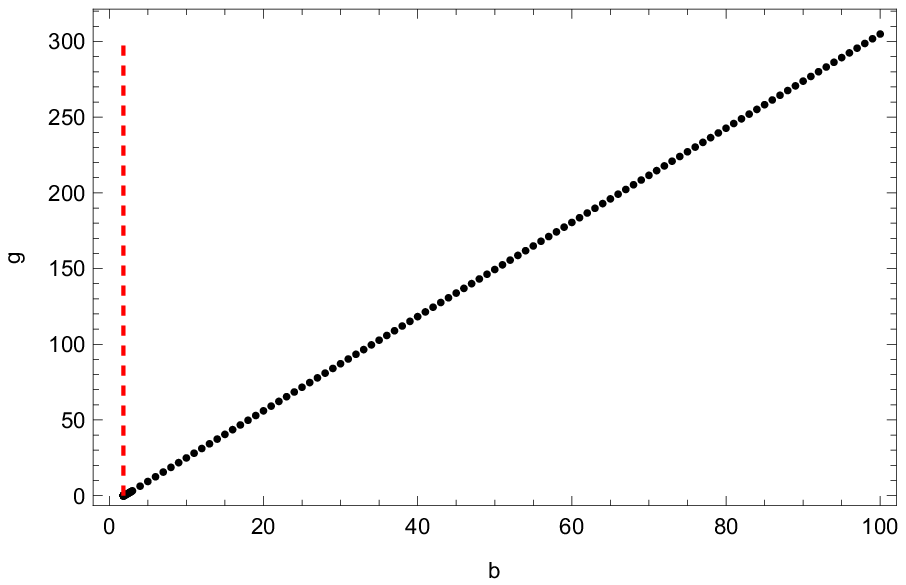}
\includegraphics[width=3in]{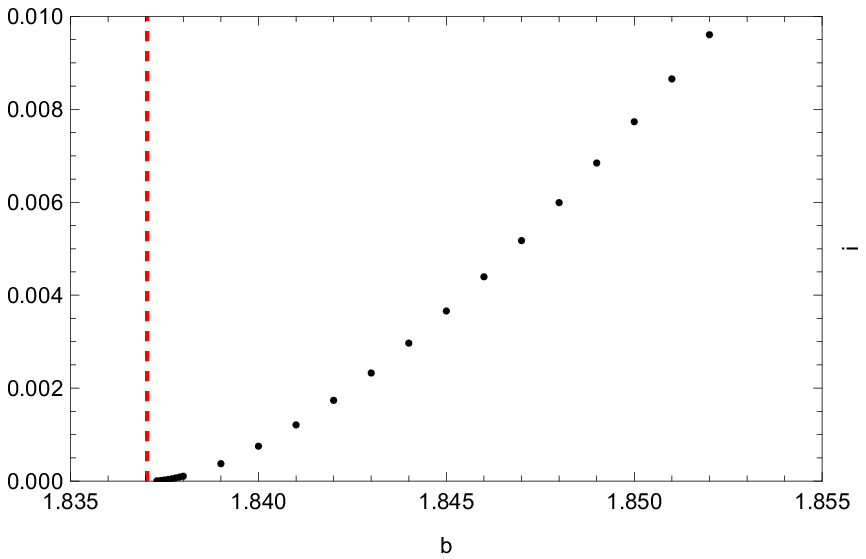}
\end{center}
  \caption{Thermal conformal order parameter $\gamma_8\equiv T^{-8}\langle\calo_8\rangle$
  of the deformed Model I diverges as $b\to b_{crit,I}+0>1$, represented by the dashed vertical red line. 
} \label{mod1g}
\end{figure}

\begin{figure}[t]
\begin{center}
\psfrag{k}[cc][][1][0]{$\kappa_{I}$}
\psfrag{b}[bb][][1][0]{$b$}
\psfrag{i}[tt][][1][0]{$\kappa_{I}$}
\includegraphics[width=3in]{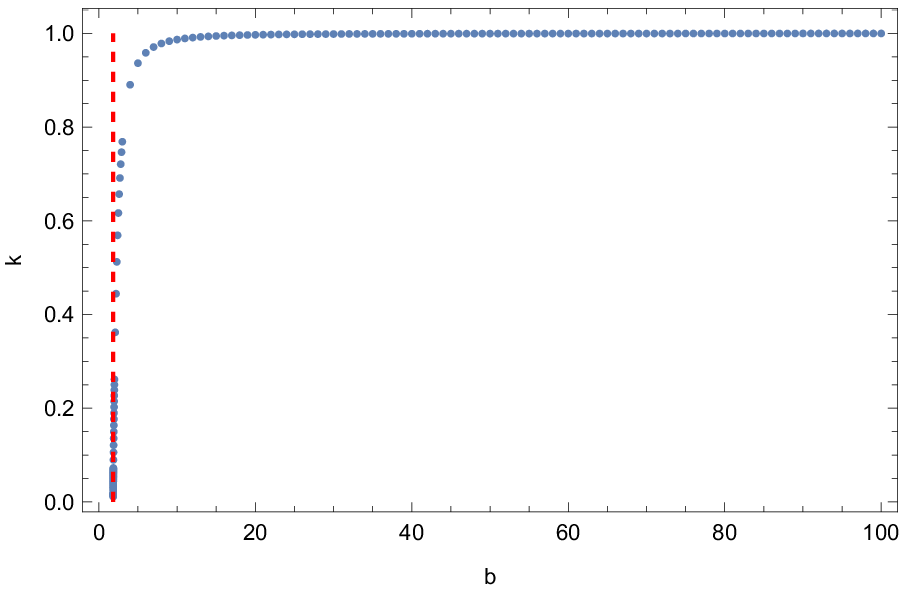}
\includegraphics[width=3in]{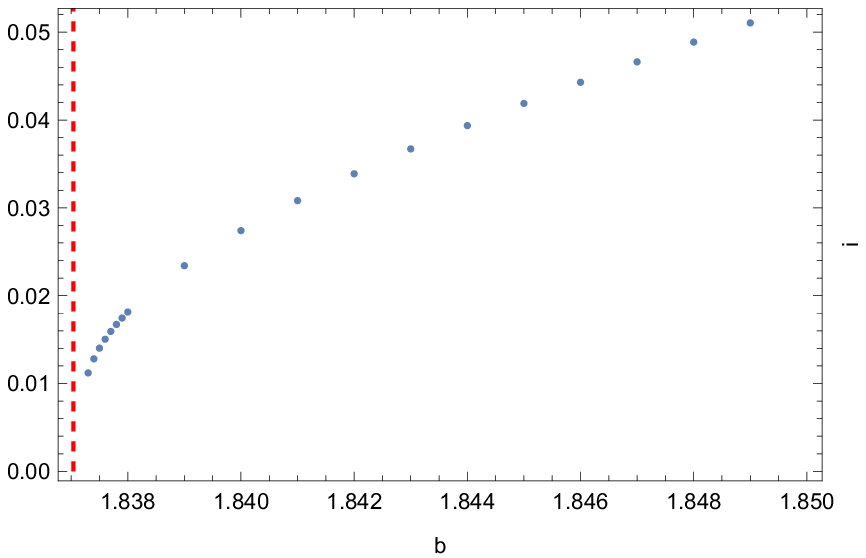}
\end{center}
  \caption{The coefficient $\kappa_I$, see \eqref{phd}, of the thermal order
  equation of state in the deformed Model I. The vertical dashed red line 
  denotes $b=b_{crit,I}$.} \label{mod1k}
\end{figure}

In the limit $b\to +\infty$ we find, see appendix \ref{pertorder},
\begin{equation}
f= \frac{9x^2(2-x)^2}{28(1+(1-x)^3)}\ \frac1b+\calo\left(b^{-2}\right)\,,\qquad h=\calo(b^{-2})\,,\qquad g=1+\calo(b^{-2})\,.
\eqlabel{mod1f1}
\end{equation}
Finite $b$ results are obtained solving \eqref{eq1mod1}-\eqref{eq3mod1}, subject to the
asymptotics \eqref{mod1bc1} and \eqref{mod1bc2}. In fig.~\ref{mod1g} we plot the 
expectation value of the order parameter, see \eqref{phd} and \eqref{data11},\eqref{kg1}.
 In fig.~\ref{mod1k} we plot the coefficient $\kappa$ of the thermal
 order equation of state in Model I as a function of the
 deformation parameter $b$, see \eqref{phd} and \eqref{kg1}. 
Notice that the order parameter diverges as $b\to b_{crit,I}$,
\begin{equation}
b_{crit,I}=1.8370(4)\,,
\eqlabel{bcriti}
\end{equation}
from above. Since $b_{crit,I}>1$, there is no conformal order in universal top-down holographic
Model I, representing a CFT dual to type IIB supergravity with the self-dual five-form flux
on a Sasaki-Einstein manifold with a breathing mode, dual to an order parameter of dimension
$\Delta=8$. Furthermore, since $\kappa_I(b)<1$, the conformal order in deformed Model I
is subdominant in canonical and microcanonical ensembles. Following \cite{Buchel:2020jfs} we expect
that this conformal order is perturbatively unstable.

\subsection{Conformal order in Model II}\label{modelii}

Conformal Model II realizes holographic dual to the KW gauge theory with a pair of order parameters:
a dimension $\Delta=8$ operator $\calo_8$ and a dimension $\Delta=6$ operator $\calo_6$,
represented by the bulk scalars  $f$ and $w$ correspondingly.
Since these scalars represent the warping of the $T^{1,1}$ base, 
along with the squashing of the $U(1)$ fibration  of the  K\"ahler-Einstein base of $T^{1,1}$,
this model is universal as well:  any holographic duality
between a CFT$ _4$ and a type IIB supergravity on AdS$ _5\times$$\tilde{Y}_{ _5}$, where
$\tilde{Y}_5$ is a five-dimensional squashed Sasaki-Einstein manifold,
can be truncated to our Model II \cite{Cassani:2010uw}.
In the latter case, the scalars $f$ and $w$ represent the breathing and the squashing modes
of $\tilde{Y}_5$.

We follow appendix \ref{pertorder} to construct conformal order in Model II.
From \eqref{eaco} we find
\begin{equation}
\begin{split}
\calp_{II}[f,w]= \calp[f,w,0]=&-12+\left(\frac{1280}{3} f^2+240 w^2\right)+\biggl(-\frac{71680}{27}f^3
-1280 f w^2\\&-1120w^3
\biggr)+\calo(\{f,w\}^4)\,,
\end{split}
\eqlabel{p2}
\end{equation}
\ie the leading nonlinear terms are $n=3$ (see \eqref{ar4}) and the unbounded direction is
along $\{f,w\}\to +\infty$. 
The $b$-deformed scalar potential takes the form,
\begin{equation}
\calp_{II}^b[f,w]=-12+\biggl(\frac{1280}{3} f^2+240w^2\biggr)+b\biggl(\calp[f,w,0]+12-\frac{1280}{3} f^2
-240 w^2\biggr)\,.
\eqlabel{p2b}
\end{equation}

\begin{figure}[t]
\begin{center}
\psfrag{u}[cc][][1][0]{$f_{1,4}$}
\psfrag{s}[bb][][1][0]{$s$}
\psfrag{i}[tt][][1][0]{$f_{1,0}^h$}
\includegraphics[width=3in]{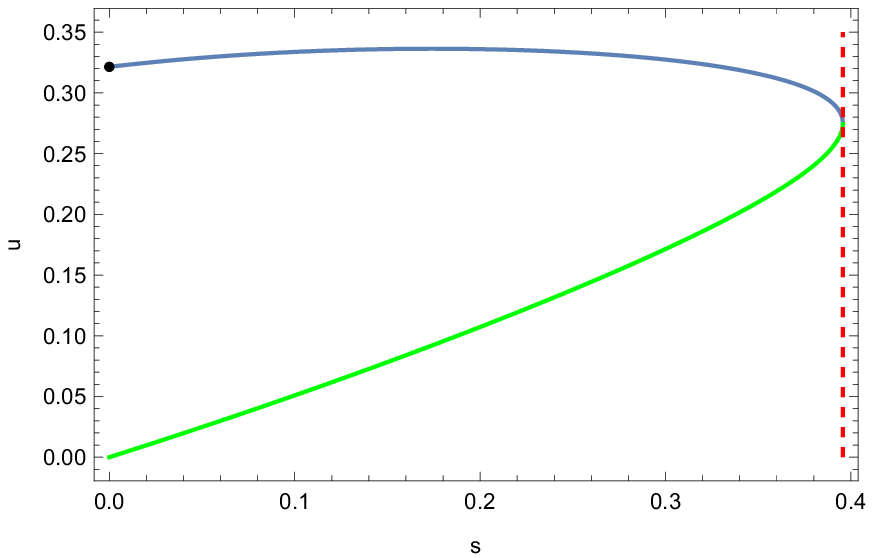}
\includegraphics[width=3in]{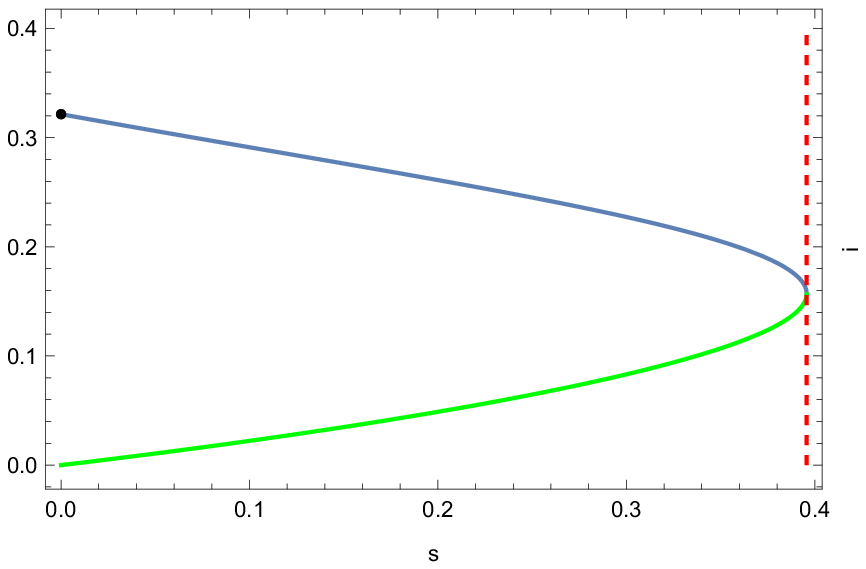}
\end{center}
  \caption{Coefficients $\{f_{1,4},f_{1,0}^h\}$, see \eqref{mod2bc1} and \eqref{mod2bc2},
of the perturbative conformal order in $b\to +\infty$ deformation of Model II, with
a parameter $s$ characterizing the cubic coupling between the scalars $f$ and $g$, see
\eqref{defmod2}.
The black dot indicates $s=0$ results reported in appendix \ref{pertorder}. There are
two branches of the conformal order for $s<s_{max}$, represented by the vertical dashed red lines.
Since $s_{max}<1$, there is no perturbative conformal order in Model II.
}
  \label{mod2fui}
\end{figure}

\begin{figure}[t]
\begin{center}
\psfrag{u}[cc][][1][0]{$w_{1,3}$}
\psfrag{s}[bb][][1][0]{$s$}
\psfrag{i}[tt][][1][0]{$w_{1,0}^h$}
\includegraphics[width=3in]{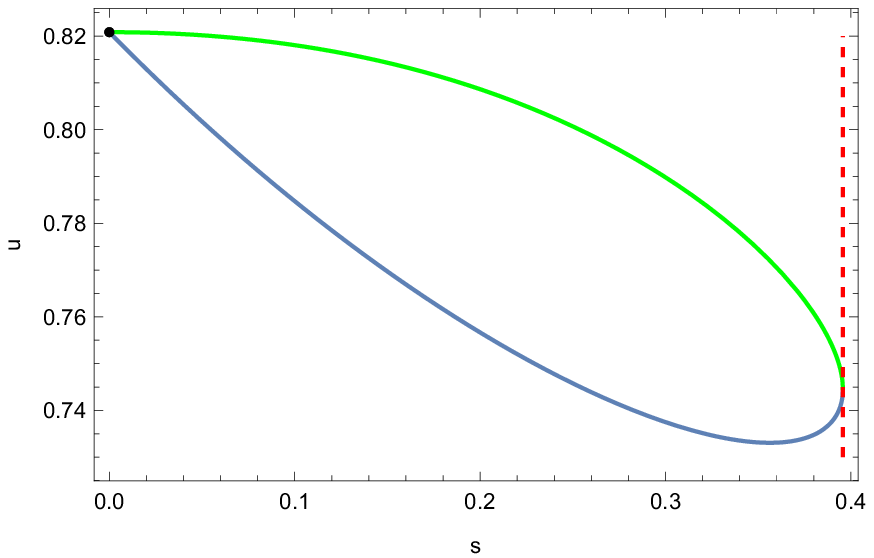}
\includegraphics[width=3in]{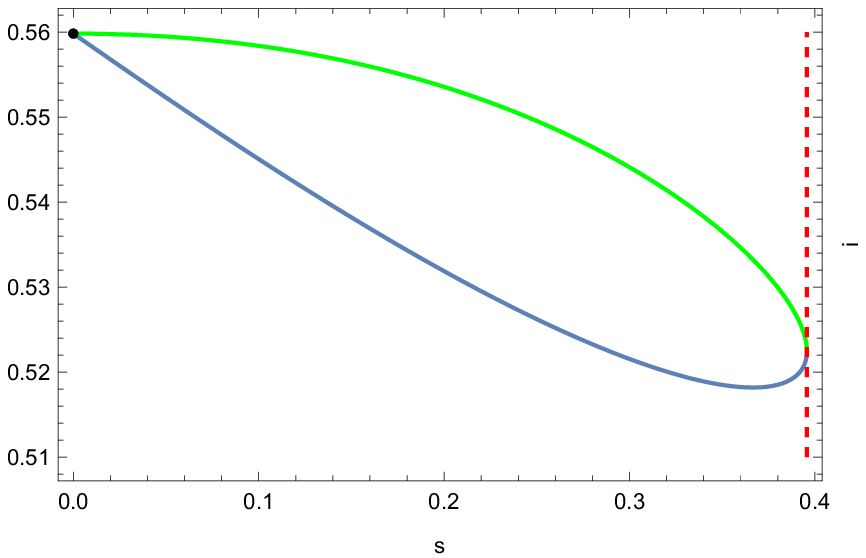}
\end{center}
  \caption{Coefficients $\{w_{1,3},w_{1,0}^h\}$, see \eqref{mod2bc1} and \eqref{mod2bc2},
of the perturbative conformal order in $b\to +\infty$ deformation of Model II, with
a parameter $s$ characterizing the cubic coupling between the scalars $f$ and $g$, see
\eqref{defmod2}.
The black dot indicates $s=0$ results reported in appendix \ref{pertorder}. There are
two branches of the conformal order for $s<s_{max}$, represented by the vertical dashed red lines.
Since $s_{max}<1$, there is no perturbative conformal order in Model II.} \label{mod2wui}
\end{figure}

Repeating the analysis as in section \ref{modeli}, in the limit $b\to +\infty$, we find
\begin{equation}
f=\frac 1b\ f_1+\calo(b^{-2})\,,\qquad  w=\frac 1b\ w_1+\calo(b^{-2})\,,\qquad
h=\calo(b^{-2})\,,\qquad  g=1+\calo(b^{-2})\,,
\eqlabel{mod2pert}
\end{equation}
where
\begin{equation}
0=f_1''+\frac{f_1'}{x-1}+\frac{4\ (56 f_1^2+9 s\ w_1^2-6 f_1)}{3\ x^2 (x-2)^2}\,,
\eqlabel{mod2f1}
\end{equation}
\begin{equation}
0=w_1''+\frac{w_1'}{x-1}+\frac{w_1\ (16 s\ f_1+21 w_1-3)}{x^2(x-2)^2}\,.
\eqlabel{mod2w1}
\end{equation}
Notice that we modified the leading nonlinear interactions in \eqref{p2} as
\begin{equation}
\biggl(-\frac{71680}{27}f^3
-1280 f w^2-1120w^3
\biggr)\ \to\ \biggl(-\frac{71680}{27}f^3
-1280s\  f w^2-1120w^3
\biggr)\,,
\eqlabel{defmod2}
\end{equation}
where a constant parameter $s$ dials the strength of the leading nonlinear coupling
between $f_1$ and $w_1$ scalars. Ultimately, we need to set $s=1$, however, it is convenient to
start at $s=0$, so that the scalars $f_1$ and $w_1$ are decoupled and we can use the results of
appendix \ref{pertorder}, and then increase $s\to 1$ .
Eqs.~\eqref{mod2f1}-\eqref{mod2w1} are solved subject to the boundary conditions:
\nxt as $x\to 0_+$, \ie at the AdS$ _5$ boundary
\begin{equation}
f_1= f_{1,4}\ x^2 +\left(f_{1,4}-\frac 34 s\ w_{1,3}^2\right) x^3+\calo(x^4)\,,\qquad
w_1=w_{1,3}\ x^{3/2}+\frac 34 w_{1,3}\ x^{5/2}+\calo(x^3)\,,
\eqlabel{mod2bc1}
\end{equation}
\nxt as $y\equiv (1-x)\to 0_+$, \ie at the horizon
\begin{equation}
f_1=f_{1,0}^h+\calo(y^2)\,,\qquad w_1=w_{1,0}^h  +\calo(y^2)\,.
\eqlabel{mod2bc2}
\end{equation}
In figs.~\ref{mod2fui} and \ref{mod2wui} we present results
for $\{f_{1,4},w_{1,3},f_{1,0}^h,w_{1,0}^h\}$ as a function of $s$.
The black dots represent $s=0$ results\footnote{With appropriate rescaling due to different
normalization of the scalar kinetic terms in \eqref{eaco} and \eqref{ap1}.}
from appendix \ref{pertorder} for
order parameters of dimensions $\Delta=8$ and $\Delta=6$. For any $0<s< s_{max}$,
\begin{equation}
s_{max}=0.3956(7)\,,
\eqlabel{smax}
\end{equation}
represented by the vertical red dashed lines, 
we find two branched of the perturbative conformal order in deformed Model II. However, since
$s_{max}<1$, we can not construct even a perturbative conformal order in $b\to +\infty$
deformation of Model II --- recall that the latter requires $s=1$.
We conclude that there is no conformal order (at least within the deformation
framework of \cite{Buchel:2020xdk}) in universal top-down holographic Model II,
representing a CFT dual to type IIB supergravity with the self-dual five-form flux on
a squashed  Sasaki-Einstein manifold with a breathing and a squashing modes,
dual to the order parameters of dimensions $\Delta=8$ and $\Delta=6$ correspondingly.

\subsection{Conformal order in Model III}\label{modeliii}

Conformal Model III realizes holographic dual to the KW gauge theory with a triplet
of order parameters:
a dimension $\Delta=8$ operator $\calo_8$, a dimension $\Delta=6$ operator $\calo_6$ and
a dimension $\Delta=3$ operator $\calo_3$,
represented by the bulk scalars  $f$, $w$ and $\lambda$ correspondingly.
A noticeable different of Model III from the models discussed in sections \ref{modeli}
and \ref{modelii} is that the former ones represent  holographic duals to
the boundary CFTs with an unbroken $U(1)_R$ symmetry; while $\langle \calo_3\rangle\ne 0$
would spontaneously break this continuous symmetry as $U(1)\to \zet_2$.

We follow appendix \ref{pertorder} to construct conformal order in Model III.
From \eqref{eaco} we find
\begin{equation}
\begin{split}
\calp_{III}[f,w,\lambda]= \calp=&-12+\left(\frac{1280}{3} f^2+240 w^2-12\lambda^2\right)
+\biggl(-\frac{71680}{27}f^3
-1280 f w^2\\&-1120w^3+64\lambda^2 (f+6w)
\biggr)+\calo(\{f,w,\lambda\}^4)\,,
\end{split}
\eqlabel{p3}
\end{equation}
\ie the leading nonlinear terms are $n=3$ (see \eqref{ar4}).
The relevant unbounded directions of the truncated potential are less obvious to identify:
as in Model II, one would expect for $\{f,w\}$ scalars to become large-positive to
drive their  effective mass \eqref{apmeff} below the BF bound; while they would 
be expected to become large-negative for the effective mass of $\lambda$ to
dip below its effective BF bound near the horizon. As we shortly demonstrate,
conformal order in deformed Model III exists when both scalars $\{f,w\}$ are negative
at the horizon. The $b$-deformed scalar potential takes the form,
\begin{equation}
\calp_{III}^b[f,w,\lambda]=-12+\biggl(\frac{1280}{3} f^2+240w^2-12\lambda^2\biggr)
+b\biggl(\calp+12-\frac{1280}{3} f^2
-240 w^2+12\lambda^2  \biggr)\,.
\eqlabel{p3b}
\end{equation}

Using the radial coordinate as in \eqref{apx}, and introducing
\begin{equation}
c_2(x)=\frac{g(x)}{(2 x-x^2)^{1/4}}\,,
\eqlabel{mod3defc2}
\end{equation}
we obtain the following equations of motion ($ '\equiv \frac{d}{dx}$)
\begin{equation}
\begin{split}
0=&f''+\frac{f'}{x-1}+ \biggl(
-\frac12 (f')^2-\frac34 (w')^2-\frac{3}{20} (\lambda')^2+\frac{9h^2}{20}-\frac{9 (x^2-2 x+2)h}{40 x (1-x) (2-x)}\\& +\frac{9}{80 x^2 (2-x)^2}
\biggr)\frac{\frac{\del\calp_{III}^b}{\del f}}{\calp_{III}^b}\,,
\end{split}
\eqlabel{eq1mod3}
\end{equation}
\begin{equation}
\begin{split}
0=&w''+\frac{w'}{x-1}+ \biggl(
-\frac13 (f')^2-\frac12 (w')^2-\frac{1}{10} (\lambda')^2+\frac{3h^2}{10}-\frac{3 (x^2-2 x+2)h}{20 x (1-x) (2-x)}\\
&+\frac{3}{40 x^2 (2-x)^2}
\biggr)\frac{\frac{\del\calp_{III}^b}{\del w}}{\calp_{III}^b}\,,
\end{split}
\eqlabel{eq1mod3w}
\end{equation}
\begin{equation}
\begin{split}
0=&\lambda''+\frac{\lambda'}{x-1}+ \biggl(
-\frac53 (f')^2-\frac52 (w')^2-\frac{1}{2} (\lambda')^2+\frac{3h^2}{2}-\frac{3 (x^2-2 x+2)h}{4 x (1-x) (2-x)}\\
&+\frac{3}{8 x^2 (2-x)^2}
\biggr)\frac{\frac{\del\calp_{III}^b}{\del \lambda}}{\calp_{III}^b}\,,
\end{split}
\eqlabel{eq1mod3l}
\end{equation}
\begin{equation}
\begin{split}
&0=h'-4 h^2+\frac{(3 x^2-6 x+4) h}{x (1-x) (2-x)}+\frac{40}{9} (f')^2+\frac{20}{3} (w')^2+\frac{4}{3} (\lambda')^2\,,
\end{split}
\eqlabel{eq2mod3}
\end{equation}
\begin{equation}
\begin{split}
&0=g'-h\ g\,.
\end{split}
\eqlabel{eq3mod3}
\end{equation}
Eqs.~\eqref{eq1mod3}-\eqref{eq2mod3} are solved subject to the boundary conditions:
\nxt as $x\to 0_+$, \ie at the AdS$ _5$ boundary
\begin{equation}
\begin{split}
&f=-\frac{3}{25} b \lambda_1^2\ x^{3/2}+f_{4}\ x^2 +\calo(x^{5/2})\,,\qquad
w=\left( \frac{3}{10}b\lambda_1^2\ \ln x+w_3\right)x^{3/2}+\calo\left(x^{5/2}\ln x\right)\,,
\\
&\lambda=x^{1/4}\left(\lambda_1\ x^{1/2}+\frac38\lambda_1\ x^{3/2}+\calo(x^2\ln x)\right)\,, 
h=-\frac{3}{10}\lambda_1^2\ x^{1/2}-\frac38\lambda_1^2\ x^{3/2}+\calo(x^2\ln^2 x)\,,\\
&g=A
\left(1-\frac{1}{5}\lambda_1^2\ x^{3/2}-\frac{3}{20}\lambda_1^2\ x^{5/2}+\calo(x^3\ln^2 x)\right)\,,
\end{split}
\eqlabel{mod3bc1}
\end{equation}
\nxt as $y\equiv (1-x)\to 0_+$, \ie at the horizon
\begin{equation}
\begin{split}
&f=f_0^h+\calo(y^2)\,,\qquad w=w_0^h+\calo(y^2)\,,\qquad \lambda=\lambda_0^h+\calo(y^2)\,,\\
&h=h_1^h y +\calo(y^3)\,,\qquad g=A\biggl(g_0^h+\calo(y^2)\biggr)\,.
\end{split}
\eqlabel{mod3bc2}
\end{equation}
Note that in total we have eight parameters $\{f_4,f_0^h,w_3,w_0^h,\lambda_1,\lambda_0^h,h_1^h,g_0^h\}$,
along with an arbitrary constant $A$. They determine the thermal conformal order
of Model III, specifically,
\begin{equation}
\begin{split}
&s=\frac{2 c_{KW}}{\pi}\ A^3\left(g_0^h\right)^3\,,\qquad \calf=-\frac{sT}{4}\,,\\
&T^{-8}\langle\calo_8\rangle=f_4\,,\qquad T^{-6}\langle\calo_6\rangle=w_3\,,\qquad T^{-3}\langle\calo_3\rangle=\lambda_1\,,\\
&T^2=\frac{A^2 \left(g_0^h\right)^2}{\pi^2 (1-2 h^h_1)} \biggl[
\left( \frac{320}{9} \left(f_0^h\right)^2-\left(\lambda_0^h\right)^2
+20 \left(w_0^h\right)^2-1\right) (b-1)
\\&+\biggl(
2 e^{-\frac{16}{3} f_0^h-2 w_0^h} \cosh(2 \lambda_0^h)
+\frac{3}{8} e^{-\frac{16}{3} f_0^h+8 w_0^h}\left(1- \cosh(4 \lambda_0^h)\right)
-\frac23 e^{-\frac{40}{3} f_0^h}-\frac13 e^{-\frac{16}{3} f_0^h-12 w_0^h}
\biggr)\\
&\times b
\biggr]\,,
\end{split}
\eqlabel{data31}
\end{equation}
for the entropy density $s$, the free energy density $\calf$,  the temperature $T$,
and the thermal order parameters $\{\calo_8,\calo_6,\calo_3\}$.
Thus,  see \eqref{phd},
\begin{equation}
\begin{split}
&\kappa_{III}=\biggl[1-2h_1^1\biggr]^{3/2}\times\biggl[
\left( \frac{320}{9} \left(f_0^h\right)^2-\left(\lambda_0^h\right)^2
+20 \left(w_0^h\right)^2-1\right) (b-1)
\\&+\biggl(
2 e^{-\frac{16}{3} f_0^h-2 w_0^h} \cosh(2 \lambda_0^h)
+\frac{3}{8} e^{-\frac{16}{3} f_0^h+8 w_0^h}\left(1- \cosh(4 \lambda_0^h)\right)
-\frac23 e^{-\frac{40}{3} f_0^h}-\frac13 e^{-\frac{16}{3} f_0^h-12 w_0^h}
\biggr)\\
&\times b
\biggr]^{-3/2}\,,\ \  \gamma_{III,\{8,6,3\}}=\{f_4,w_3,\lambda_1\}\,,
\end{split}
\eqlabel{kg3}
\end{equation}
where the subscript $ _{III}$ refers to Model III.

\begin{figure}[t]
\begin{center}
\psfrag{f}[cc][][1][0]{$f_0^h$}
\psfrag{b}[bb][][1][0]{$b$}
\psfrag{w}[tt][][1][0]{$w_0^h$}
\includegraphics[width=3in]{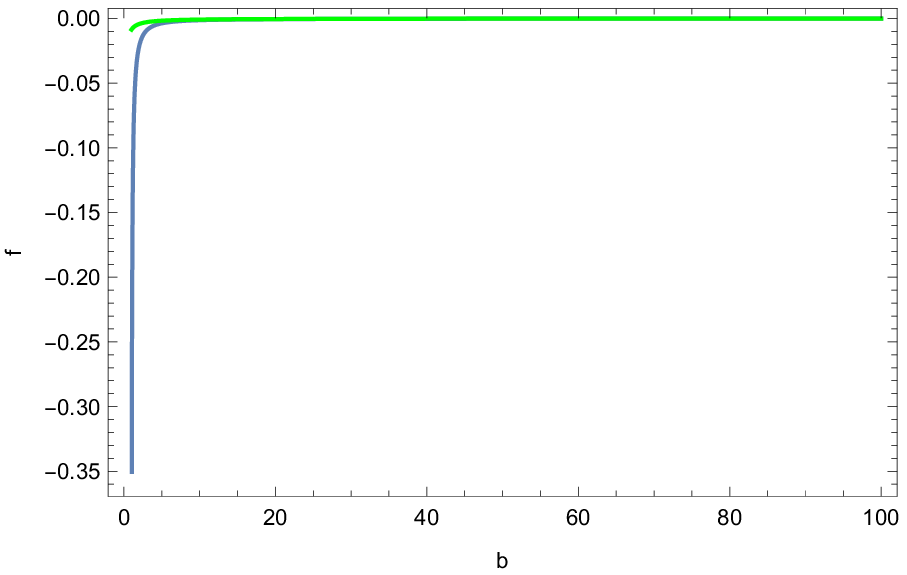}
\includegraphics[width=3in]{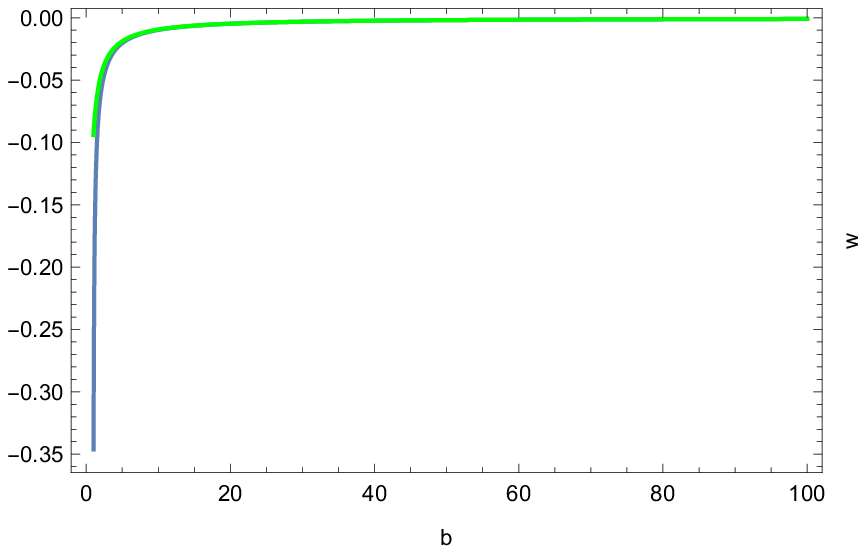}
\end{center}
  \caption{Conformal order in the deformed Model III arises due to negative
  values \eqref{mod3bc2} of $f$ and $w$ scalars at the
  Schwarzschild horizon (blue curves), resulting in the 
  effective mass of $\lambda$-scalar, see \eqref{meffla}, below the
  effective BF bound. The green curves represent
  large-$b$ approximations to $\{f_0^h,w_0^h\}$, see \eqref{normmod3}.} \label{fwhor}
\end{figure}

\begin{figure}[t]
\begin{center}
\psfrag{m}[cc][][1][0]{$m_{eff,\lambda}^2$}
\psfrag{b}[bb][][1][0]{$b$}
\psfrag{w}[tt][][1][0]{$w_0^h$}
\includegraphics[width=4in]{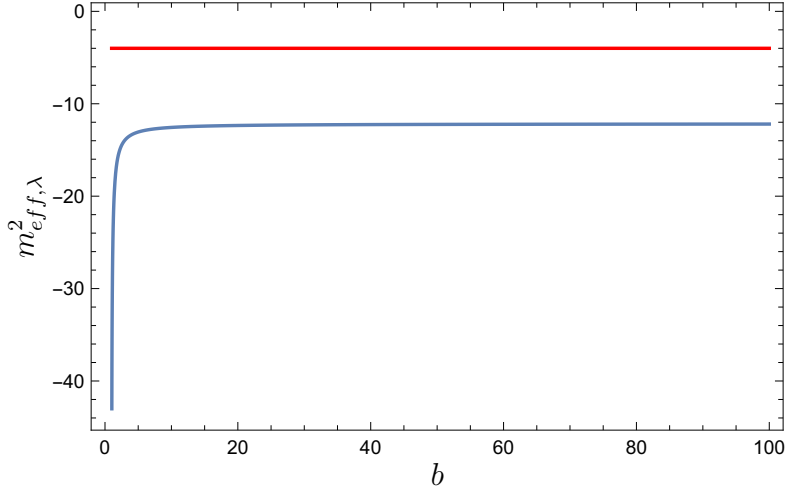}
\end{center}
  \caption{Effective mass of the $\lambda$-scalar evaluated at the
  Schwarzschild horizon (blue curve), see \eqref{meffla}. The red line represents AdS$ _5$
  BF bound.} \label{meff2}
\end{figure}

In the limit $b\to +\infty$ we find, see appendix \ref{pertorder},
\begin{equation}
\{f,w,\lambda\}= \{f_1,w_1,\ell_1\}\ \frac1b+\calo\left(b^{-2}\right)\,,\qquad h=\calo(b^{-2})\,,\qquad g=1+\calo(b^{-2})\,,
\eqlabel{mod3f1}
\end{equation}
with
\begin{equation}
\begin{split}
&f_{4}=0.27(3)\frac 1b+\calo(b^{-2})\,,\ f_0^h=-0.008(6)\frac 1b+\calo(b^{-2})\,,\
w_3=0.16(5)\frac 1b+\calo(b^{-2})\,,\\
&w_0^h=-0.094(0)\frac 1b+\calo(b^{-2})\,,\ \lambda_1=0.95(2)\frac 1b+\calo(b^{-2})\,,\
\lambda_0^h=0.59(3)\frac 1b+\calo(b^{-2})\,.
\end{split}
\eqlabel{normmod3}
\end{equation}
Finite $b$ results are obtained solving \eqref{eq1mod3}-\eqref{eq3mod3}, subject to the
asymptotics \eqref{mod3bc1} and \eqref{mod3bc2}.
Notice that both $f_0^h$ and $w_0^h$ are negative as $b\gg 1$; as fig.~\ref{fwhor} shows
they continue to stay negative for finite $b$, driving the effective mass of $\lambda$
at the Schwarzschild horizon below the effective BF bound,
\begin{equation}
\begin{split}
m_{eff,\lambda}^2\bigg|_{horizon}=&\Delta(\Delta-4)+16b(f+6w)\bigg|_{horizon}^{\Delta=3}=-3 +
16b\left(f_0^h+6w_0^h\right)\\
=&-12.16(1)+\calo(b^{-1})\,,
\end{split}
\eqlabel{meffla}
\end{equation}
and causing the
condensation of all the scalars, see fig.~\ref{meff2}.

\begin{figure}[t]
\begin{center}
\psfrag{g}[cc][][1][0]{$1/\gamma_{I,8}$}
\psfrag{b}[bb][][1][0]{$b$}
\psfrag{i}[tt][][1][0]{$1/\gamma_{I,8}$}
\includegraphics[width=3in]{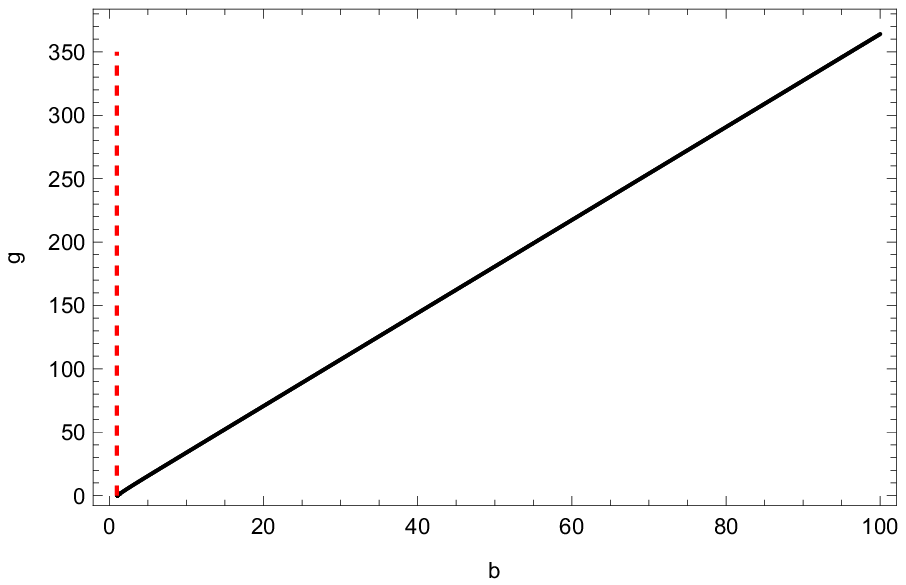}
\includegraphics[width=3in]{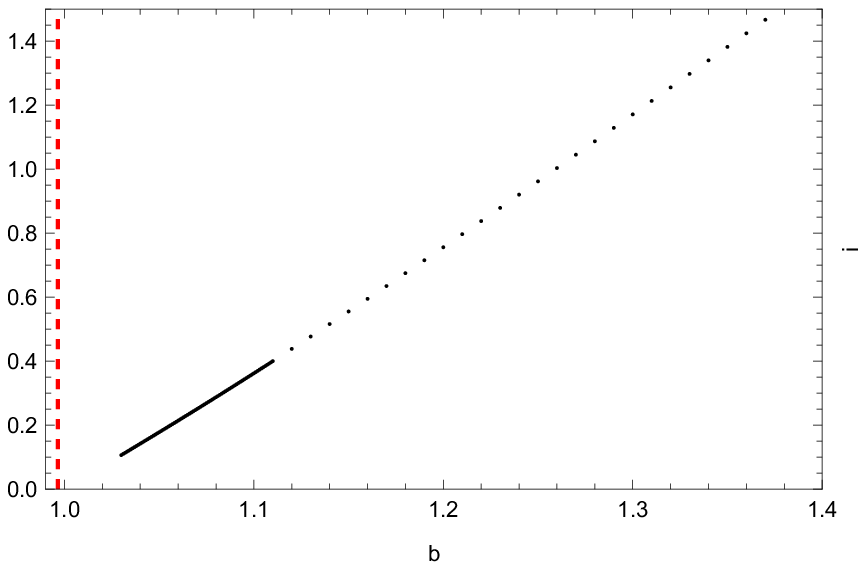}
\end{center}
  \caption{Thermal conformal order parameter $\gamma_8\equiv T^{-8}\langle\calo_8\rangle$
  of the deformed Model III diverges as $b\to b_{crit,III}+0$, represented by the dashed vertical red line. 
} \label{mod3g8}
\end{figure}

\begin{figure}[t]
\begin{center}
\psfrag{g}[cc][][1][0]{$1/\gamma_{I,6}$}
\psfrag{b}[bb][][1][0]{$b$}
\psfrag{i}[tt][][1][0]{$1/\gamma_{I,6}$}
\includegraphics[width=3in]{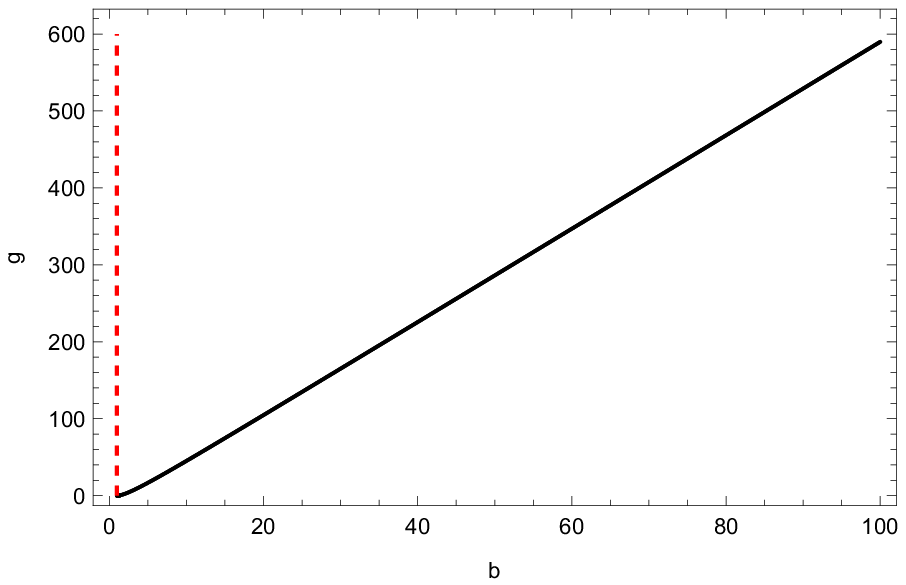}
\includegraphics[width=3in]{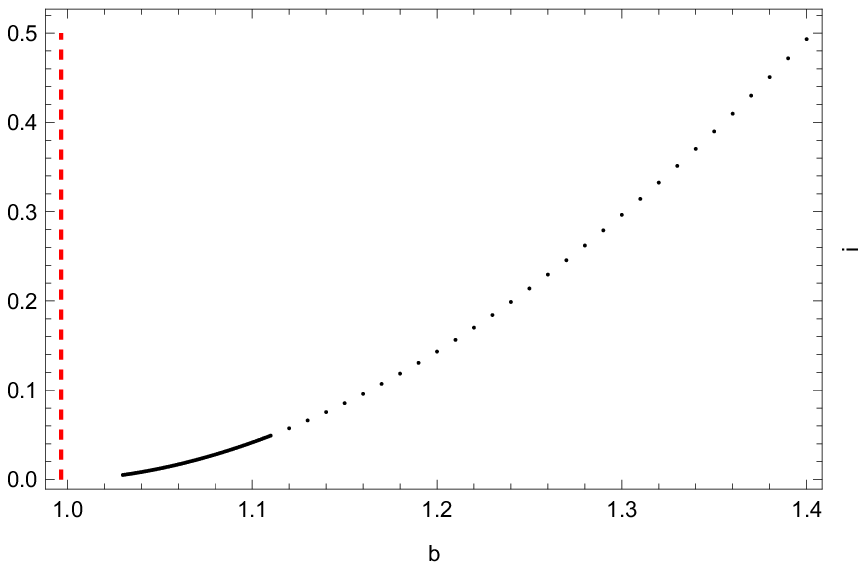}
\end{center}
  \caption{Thermal conformal order parameter $\gamma_6\equiv T^{-6}\langle\calo_6\rangle$
  of the deformed Model III diverges as $b\to b_{crit,III}+0$, represented by the dashed vertical red line. 
} \label{mod3g6}
\end{figure}

\begin{figure}[t]
\begin{center}
\psfrag{g}[cc][][1][0]{$1/\gamma_{I,3}$}
\psfrag{b}[bb][][1][0]{$b$}
\psfrag{i}[tt][][1][0]{$1/\gamma_{I,3}$}
\includegraphics[width=3in]{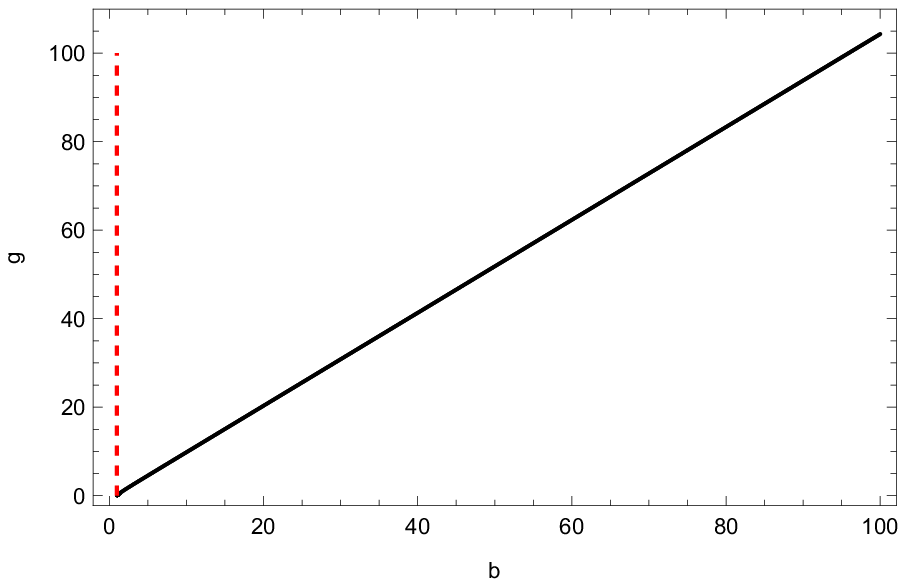}
\includegraphics[width=3in]{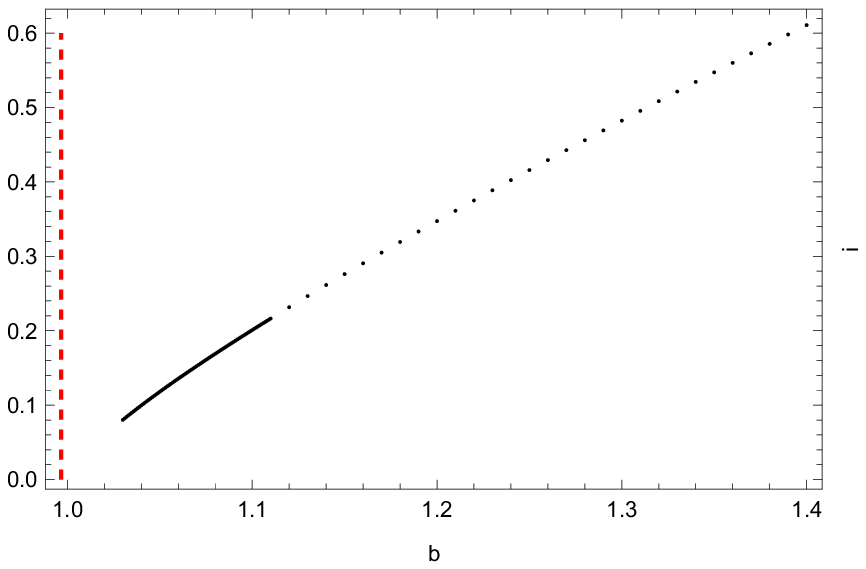}
\end{center}
  \caption{Thermal conformal order parameter $\gamma_3\equiv T^{-3}\langle\calo_3\rangle$
  of the deformed Model III diverges as $b\to b_{crit,III}+0$, represented by the dashed vertical red line. 
} \label{mod3g3}
\end{figure}

\begin{figure}[t]
\begin{center}
\psfrag{k}[cc][][1][0]{$\kappa_{III}$}
\psfrag{b}[bb][][1][0]{$b$}
\psfrag{i}[tt][][1][0]{$\kappa_{III}$}
\includegraphics[width=3in]{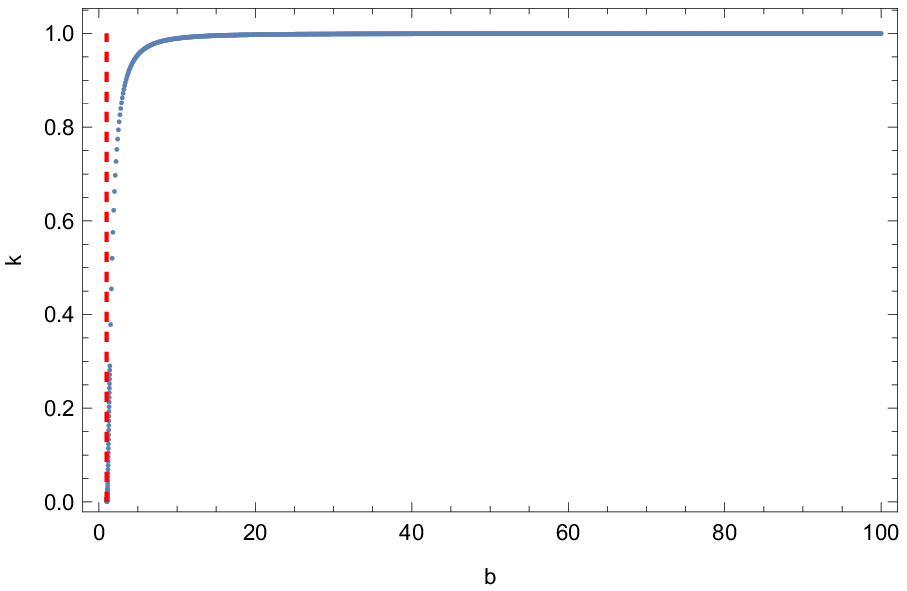}
\includegraphics[width=3in]{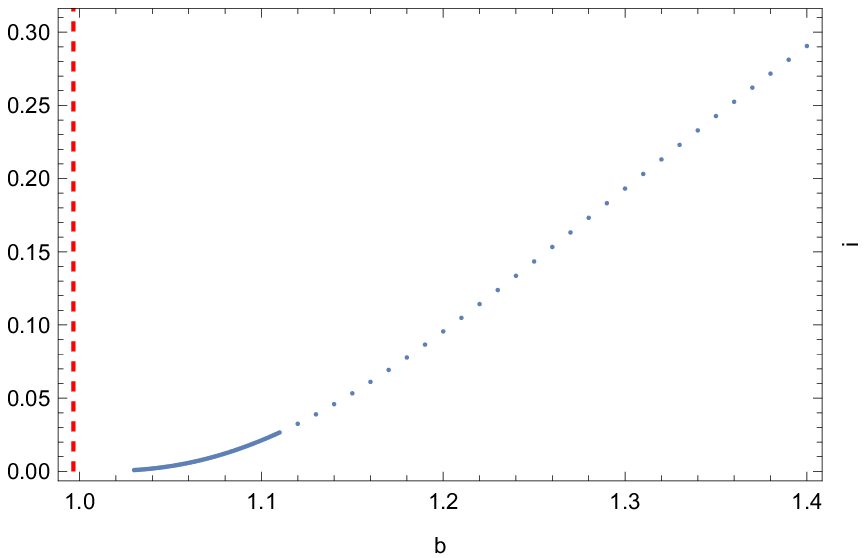}
\end{center}
  \caption{The coefficient $\kappa_{III}$, see \eqref{phd}, of the thermal order
  equation of state in the deformed Model III. The vertical dashed red line 
  denotes $b=b_{crit,III}$.} \label{mod3k}
\end{figure}

In figs.~\ref{mod3g8}-\ref{mod3g3} we plot the 
expectation values of the order parameters $\gamma_{III,\{8,6,3\}}$, see \eqref{phd} and \eqref{data11},\eqref{kg3}.
 In fig.~\ref{mod3k} we plot the coefficient $\kappa$ of the thermal
 order equation of state in Model III as a function of the
 deformation parameter $b$, see \eqref{phd} and \eqref{kg3}. 
We obtained reliable numerical results for $b \gtrsim 1.03$; their extrapolation  
to smaller values of $b$ predicts that all the order parameters diverge as $b\to b_{crit,III}$,
\begin{equation}
b_{crit,III}=0.99(7)\bigg|_{\rm extrapolation}\,,
\eqlabel{bcritiii}
\end{equation}
from above. We take closeness of $b_{crit,III}$ to $1$ as the strong indication
that the actual value of $b_{crit,III}$ is precisely $1$. We thus conclude that
while there is a conformal order in $b$-deformed Model III, it disappears
once this model becomes a realization of the top-down holography, much like in top-down holographic
models discussed in \cite{Buchel:2020xdk,Buchel:2021ead}.
Since $\kappa_{III}(b)<1$, the conformal order in deformed Model III
is subdominant in canonical and microcanonical ensembles. Following \cite{Buchel:2020jfs} we expect
that this conformal order is perturbatively unstable.

\section{Conclusion}\label{conclude}

In this paper we searched for the holographic conformal order \cite{Buchel:2020thm} on
warped deformed conifold with fluxes \cite{Buchel:2021yay} --- representing the supergravity
dual to Klebanov-Strassler cascading gauge theory \cite{Klebanov:2000hb}. 
We focused on two exotic phases \cite{Buchel:2009ge} of the black branes on the conifold:
\begin{itemize}
\item the Klebanov-Strassler black branes \cite{Buchel:2018bzp}, realizing the spontaneous
breaking of the continuous $R$-symmetry;
\item the branch of Klebanov-Tseytlin black branes with the negative specific heat
\cite{Buchel:2009bh}.
\end{itemize}
The two classes of the black branes have an intrinsic scale, the holographic
dual to the strong coupling
scale $\Lambda$ of the boundary cascading gauge theory. Thus, to find the conformal
order one needs to construct these black branes in the high temperature limit
$\frac T\Lambda\to 0$. We showed that as one increases the temperature, these
exotic black branes on the conifold become ever more stringy. As a result, the conformal order
can not be reached within the controllable supergravity approximation.

In the second part of the paper we changed the strategy: instead of starting with the
non-conformal exotic black branes and taking the high-temperature limit, we
took the conformal limit in the effective action of type IIB supergravity
on warped deformed conifold with fluxes \cite{Buchel:2010wp}. The latter
limit effectively removes the fractional D3 branes from the model.
The resulting five-dimensional effective action has seven bulk scalars; but only
three of these scalars have the nonlinear potential. The nonlinearity of the
scalar potential appears to be crucial in holographic
examples of the conformal order constructed so far
\cite{Buchel:2009ge,Buchel:2020thm,Buchel:2020xdk,Buchel:2021ead}. Taking this fact as
a hint, we consistently truncated the ``conformal'' effective action to
these three scalars. Within this action, we further identified
two additional universal consistent truncations:
\begin{itemize}
\item type IIB supergravity on warped Sasaki-Einstein manifolds with the self-dual
five-form flux \cite{Buchel:2009bh,Cassani:2010uw};
\item type IIB supergravity on warped and squashed Sasaki-Einstein manifolds with the self-dual
five-form flux \cite{Buchel:2009bh,Cassani:2010uw}.
\end{itemize}
We showed that in both truncations one can construct the conformal order,
provided one deforms the bulk scalar potential obtained from the Kaluza-Klein
reduction on the Sasaki-Einstein manifold. However, before the deformation parameter is
removed, the conformal order disappears.
The same story repeats in the full  effective action on the
warped deformed conifold with the three bulk scalars --- the conformal order exists, as long as
the scalar potential is (arbitrarily small) deformed from the
one obtained from the type IIB supergravity Kaluza-Klein reduction.  
Along with the previous results \cite{Buchel:2020xdk,Buchel:2021ead}
we believe this sends a powerful message:
there is a 'real' holography and a 'toy' one; conformal order is possible only
in the latter. This begs a question: how much one can trust phenomenological
holographic models to predict String Theory Universe phenomena?

\section*{Acknowledgments}
This research is supported in part by Perimeter Institute for
Theoretical Physics.  Research at Perimeter Institute is supported in
part by the Government of Canada through the Department of Innovation,
Science and Economic Development Canada and by the Province of Ontario
through the Ministry of Colleges and Universities. This work was
further supported by NSERC through the Discovery Grants program.

\appendix

\section{Perturbative holographic thermal conformal order}
\label{pertorder}

Consider thermal conformal order in the phenomenological model 
\begin{equation}
S_5=\frac{c}{2\pi^2} \int_{\calm_{5}} \vol_{\calm_5} \biggl\{R-\frac 12 (\nabla\phi)^2-\calp^b[\phi]
\biggr\}\,,
\eqlabel{ap1}
\end{equation}
with the scalar potential $\calp^b[\phi]$  given by \eqref{ar5}, \ie
\begin{equation}
\calp^b=-12 +\frac 12 \Delta(\Delta-4) \phi^2-b \sum_{k=n\ge 3}^\infty p_k\ \phi^k\,,
\eqlabel{ap2}
\end{equation}
where $p_n> 0$, and the following next nonzero coefficient $p_k$ is for $k=n+p\ge n+1$.
We assume $\Delta \ge 2$. To this end, we assume the black brane metric ansatz
\begin{equation}
ds_5=^2=-c_1^2\ dt^2+ c_2^2\ d\bm{x}^2+c_3^2\ dr^2\,,\qquad c_i=c_i(r)\,,
\eqlabel{ap3}
\end{equation}
along with $\phi=\phi(r)$, all depending on the radial coordinate $r\in [r_0,+\infty)$.
There is the smooth Schwarzschild horizon as $r\to r_{0_+}$, \ie
\begin{equation}
\lim_{r\to r_{0_+}} c_1^2 =0\,,
\eqlabel{ap4}
\end{equation}
and asymptotic AdS$ _5$ geometry, \ie as $r\to \infty$,
\begin{equation}
c_1^2\to r^2\,,\quad c_2\to r^2\,,\quad c_3^2\to r^{-2}\,,\quad \phi\to \frac{\langle\calo_\Delta\rangle}{r^\Delta} \,,
\eqlabel{ap5}
\end{equation}
where $\langle\calo_\Delta\rangle\ne 0$ is the order parameter, see \eqref{phd}.  

From the effective action \eqref{ap1} we derive the following equations of motion
\begin{equation}
0=c_1''+c_1'\ \left[\ln \frac{c_2^3}{c_3}\right]'+\frac 13 c_1 c_3^2\ \calp^b\,,
\eqlabel{apeom1}
\end{equation}
\begin{equation}
0=c_2''+c_2'\ \left[\ln \frac{c_1c_2^2}{c_3}\right]'+\frac 13 c_2 c_3^2\ \calp^b\,,
\eqlabel{apeom2}
\end{equation}
\begin{equation}
0=(\phi')^2-12 \left[\ln c_2\right]' \left[\ln (c_1 c_2)\right]'-2 c_3^2\ \calp^b\,,
\eqlabel{apeom3}
\end{equation}
\begin{equation}
0=\phi''+\phi'\ \left[\ln\frac{c_1c_2^3}{c_3}\right]'-c_3^2\ \frac{\del \calp^b}{\del\phi}\,,
\eqlabel{apeom4}
\end{equation}
where $' \equiv \frac{d}{dr}$.
Equations \eqref{apeom1}-\eqref{apeom4} can be systematically analyzed perturbatively as $b\to +\infty$.
Introducing a new radial coordinate \cite{Buchel:2003ah},
\begin{equation}
1-x=\frac{c_1}{c_2}\,,\qquad x\in(0,1]\,,
\eqlabel{apx}
\end{equation}
with $x\to 1_-$ representing the regular Schwarzschild horizon \eqref{ap4}, and $x\to 0_+$
the boundary asymptotes \eqref{ap5}, we find  
\begin{equation}
\begin{split}
&c_1=(1-x) c_2\,,\qquad c_2=\frac{\pi T}{(2x-x^2)^{1/4}}\biggl[1+\calo\left(\left(\frac 1b\right)^{\frac{2}{n-2}}\right)\biggr]\,,
\\
&c_3^2\ dr^2=\frac{dx^2}{4(2x-x^2)^2}  \biggl[1+\calo\left(\left(\frac 1b\right)^{\frac{2}{n-2}}\right)\biggr]\,,
\\
&\phi=\left(\frac{1}{2 p_n b}\right)^{\frac{1}{n-2}}\biggl[\phi_1(x)
+\calo\left(\left(\frac 1b\right)^{\frac{2}{n-2}},\left(\frac 1b\right)^{\frac{p}{n-2}}\right)\biggr]\,,
\end{split}
\eqlabel{apmetric}
\end{equation}
where $T$ is the Hawking temperature of the horizon, and the leading as $b\to +\infty$
scalar hair $\phi_1$ satisfies 
\begin{equation}
0=\phi_1''+\frac{\phi_1'}{x-1}+\frac{\phi_1(n\phi_1-2 \Delta(\Delta-4))}{8x^2(2-x)^2}\,,
\eqlabel{apscalar}
\end{equation}
subject to the boundary conditions:
\nxt as $x\to 0_+$, \ie at the AdS$ _5$ boundary
\begin{equation}
\phi_1\to f_{1,0}\ x^{\Delta/4}\qquad \Longrightarrow\qquad
T^{-\Delta}\langle\calo_\Delta\rangle\ \propto\ f_{1,0}\ \left(\frac 1b\right)^{\frac{1}{n-2}} \,,
\eqlabel{apbc1}
\end{equation}
\nxt as $x\to 1_-$, \ie at the horizon
\begin{equation}
\phi_1\to f_{1,h}\,.
\eqlabel{apbc2}
\end{equation}
Once the perturbative (as $b\to +\infty$) solution \eqref{apmetric} is constructed,
the finite-$b$ conformal order can be analyzed numerically, incrementally decreasing the
deformation parameter $b$ \cite{Buchel:2020thm,Buchel:2020xdk,Buchel:2021ead}.

The reason why we expect the specific large-$b$ scaling of the conformal order as in
\eqref{apmetric} was given in  \cite{Buchel:2020xdk}:
\begin{itemize}
\item Recall the story of the holographic superconductor \cite{Gubser:2005ih,Hartnoll:2008kx}.
A scalar field in asymptotically AdS geometry must have a mass above the (space-time dependent)
Breitenlohner-Freedman (BF) bound to avoid the condensation. In the vicinity
of the Schwarzschild horizon, the  BF bound can be modified either by
changing the effective dimensionality of the space-time (as in the extremal limit
of a Reissner-Nordstrom black brane) \cite{Hartnoll:2008kx}, or via nonlinear scalar
interactions, leading to large negative contribution of the effective mass \cite{Gubser:2005ih}.
Thus, it is possible for a scalar field to be {\it above} the BF bound close to the
AdS boundary, and {\it below} the effective BF bound close to the horizon. This scenario triggers
the condensation of the scalar - the black brane develops the scalar hair. Close to the
transition point, the condensation can be studied in the probe approximation --- neglecting
the backreaction.
\item The  conformal order realizes the black brane horizon scalarization
mechanism of \cite{Gubser:2005ih}. It becomes a {\it probe approximation} in the
limit $b\to +\infty$. Indeed, with the scaling 
\begin{equation}
\phi\ \propto\ \left(\frac 1b\right)^{\frac{1}{n-2}}\,,\qquad n\ge 3\,,
\eqlabel{aps1}
\end{equation}
the scalar backreaction on the AdS$ _5$-Schwarzschild geometry vanishes as in \eqref{apmetric}.
Furthermore, an order $k\ge n$ monomial in the scalar potential \eqref{ap2}
scales as
\begin{equation}
b\ p_k\ \phi^k \ \propto\ p_k\ \left(\frac 1b\right)^{\frac{2}{n-2}+\frac{k-n}{n-2}}\,,
\eqlabel{aps2}
\end{equation}
while the mass term in the scalar potential \eqref{ap2}, as well as the kinetic term, scale as
\begin{equation}
(\nabla\phi)^2\ \sim\ \Delta(\Delta-4)\ \phi^2 \ \propto\ 
\left(\frac 1b\right)^{\frac{2}{n-2}}\,.
\eqlabel{aps3}
\end{equation}
Thus, any nonlinear term in the scalar potential with $k>n$ is subleading in the
$b\to +\infty$ limit compare to the quadratic scalar terms in the effective action \eqref{ap1},
while the leading nonlinear $k=n$ term scales precisely as the former.
\item To summarize, given \eqref{aps1}, the holographic model \eqref{ap1} represents the
probe approximation of the scalar $\phi$ on AdS$ _5$-Schwarzschild geometry
with the effective action:
\begin{equation}
\begin{split}
S_{scalar}=&\int_{\calm_5={\rm AdS}_5-{\rm Schwarzschild}} {\vol}_{\calm_5}
\biggl\{-\frac 12 (\nabla\phi)^2-\frac 12 m_{eff}^2\ \phi^2\biggr\}
\\
\propto & \int_{\calm_5={\rm AdS}_5-{\rm Schwarzschild}} \vol_{\calm_5}
\biggl\{-\frac 12 (\nabla\phi_1)^2-\frac 12 m_{eff}^2\ \phi_1^2\biggr\}\,,
\end{split}
\eqlabel{approbe}
\end{equation}
\end{itemize}
where
\begin{equation}
\begin{split}
m_{eff}^2=\Delta(\Delta-4)- 2p_n b\ \phi^{n-2}=\Delta(\Delta-4)- \phi_1^{n-2}\,,
\end{split}
\eqlabel{apmeff}
\end{equation}
and the scalar field $\phi_1$ satisfies \eqref{apscalar}. 
Notice from \eqref{apmeff}  that a large {\it positive} value of $\phi_1$ at the horizon,
see \eqref{apbc2}, can make $m_{eff}^2$ sufficiently negative, and trigger the scalarization
as in  \cite{Gubser:2005ih}. 

\begin{figure}[t]
\begin{center}
\psfrag{d}[cc][][1][0]{$\Delta$}
\psfrag{u}[bb][][1][0]{$f_{1,0}$}
\psfrag{i}[tt][][1][0]{$f_{1,h}$}
\includegraphics[width=3in]{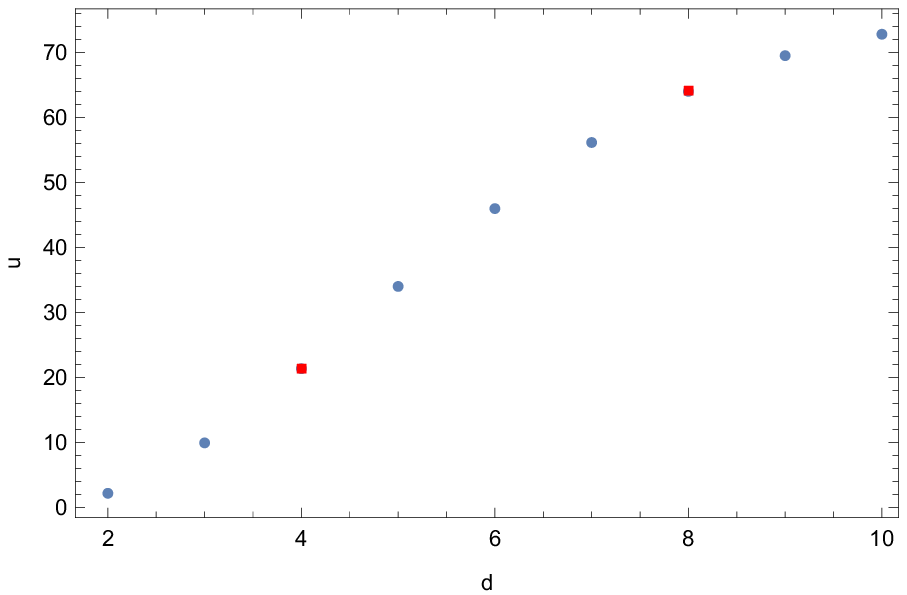}
\includegraphics[width=3in]{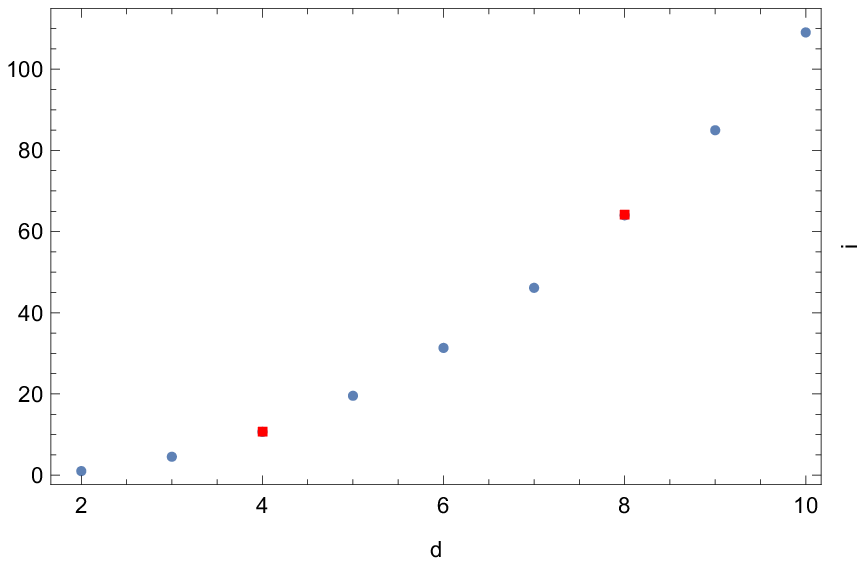}
\end{center}
  \caption{Normalizable coefficients, \eqref{apbc1} and \eqref{apbc2},
  of the bulk scalar field dual to the conformal order
  parameter $\calo_\Delta$ in the limit $b\to +\infty$. The red squares represent
  the analytical results  obtained for $\Delta=4$ and $\Delta=8$ correspondingly. 
} \label{figap1}
\end{figure}

We now return to the analysis of \eqref{apscalar}-\eqref{apbc2}. We
consider\footnote{Generalizations to other $n$ and $\Delta$ are straightforward.}
$n=3$, as this case  will be of relevance for the discussion in section \ref{ksdef},
and consider integer values of $\Delta=\{2,\cdots 10\}$. Numerical results for
$\{f_{1,0}, f_{1,h}\}$ are presented in fig.~\ref{figap1}; for $\Delta=4$ and
$\Delta=8$ \eqref{apscalar}-\eqref{apbc2} can be solved analytically:
\begin{equation}
\begin{split}
&\Delta=4:\qquad \phi_1=\frac{32}{3}\ x(2-x)\qquad \Longrightarrow\qquad \{f_{1,0},f_{1,h}\}
=\{\frac{64}{3},\frac{32}{3}\}\,,
\\
&\Delta=8:\qquad \phi_1=64\ \frac{x^2(2-x)^2}{(1+(1-x)^2)^2}\qquad \Longrightarrow\qquad \{f_{1,0},f_{1,h}\}
=\{{64},{64}\}\,.
\end{split}
\eqlabel{d4d8}
\end{equation}

\bibliographystyle{JHEP}
\bibliography{ksorder}

\providecommand{\href}[2]{#2}\begingroup\raggedright\begin{thebibliography}{10}

\bibitem{Chai:2020zgq}
N.~Chai, S.~Chaudhuri, C.~Choi, Z.~Komargodski, E.~Rabinovici and M.~Smolkin,
  \emph{{Thermal Order in Conformal Theories}},
  \href{http://dx.doi.org/10.1103/PhysRevD.102.065014}{\emph{Phys. Rev. D} {\bf
  102} (2020) 065014}, [\href{https://arxiv.org/abs/2005.03676}{{\tt
  2005.03676}}].

\bibitem{Buchel:2009ge}
A.~Buchel and C.~Pagnutti, \emph{{Exotic Hairy Black Holes}},
  \href{http://dx.doi.org/10.1016/j.nuclphysb.2009.08.017}{\emph{Nucl. Phys. B}
  {\bf 824} (2010) 85--94}, [\href{https://arxiv.org/abs/0904.1716}{{\tt
  0904.1716}}].

\bibitem{Buchel:2020thm}
A.~Buchel, \emph{{Thermal order in holographic CFTs and no-hair theorem
  violation in black branes}},
  \href{http://dx.doi.org/10.1016/j.nuclphysb.2021.115425}{\emph{Nucl. Phys. B}
  {\bf 967} (2021) 115425}, [\href{https://arxiv.org/abs/2005.07833}{{\tt
  2005.07833}}].

\bibitem{Buchel:2020jfs}
A.~Buchel, \emph{{Fate of the conformal order}},
  \href{http://dx.doi.org/10.1103/PhysRevD.103.026008}{\emph{Phys. Rev. D} {\bf
  103} (2021) 026008}, [\href{https://arxiv.org/abs/2011.11509}{{\tt
  2011.11509}}].

\bibitem{Buchel:2020xdk}
A.~Buchel, \emph{{SUGRA/Strings like to be bald}},
  \href{http://dx.doi.org/10.1016/j.physletb.2021.136111}{\emph{Phys. Lett. B}
  {\bf 814} (2021) 136111}, [\href{https://arxiv.org/abs/2007.09420}{{\tt
  2007.09420}}].

\bibitem{Chaudhuri:2020xxb}
S.~Chaudhuri, C.~Choi and E.~Rabinovici, \emph{{Thermal order in large N
  conformal gauge theories}},
  \href{http://dx.doi.org/10.1007/JHEP04(2021)203}{\emph{JHEP} {\bf 04} (2021)
  203}, [\href{https://arxiv.org/abs/2011.13981}{{\tt 2011.13981}}].

\bibitem{Chai:2021djc}
N.~Chai, A.~Dymarsky and M.~Smolkin, \emph{{A model of persistent breaking of
  discrete symmetry}},  \href{https://arxiv.org/abs/2106.09723}{{\tt
  2106.09723}}.

\bibitem{Chaudhuri:2021dsq}
S.~Chaudhuri and E.~Rabinovici, \emph{{Symmetry breaking at high temperatures
  in large N gauge theories}},  \href{https://arxiv.org/abs/2106.11323}{{\tt
  2106.11323}}.

\bibitem{Buchel:2021ead}
A.~Buchel, \emph{{Compactified holographic conformal order}},
  \href{http://dx.doi.org/10.1016/j.nuclphysb.2021.115605}{\emph{Nucl. Phys. B}
  {\bf 973} (2021) 115605}, [\href{https://arxiv.org/abs/2107.05086}{{\tt
  2107.05086}}].

\bibitem{Chai:2021tpt}
N.~Chai, A.~Dymarsky, M.~Goykhman, R.~Sinha and M.~Smolkin, \emph{{A model of
  persistent breaking of continuous symmetry}},
  \href{https://arxiv.org/abs/2111.02474}{{\tt 2111.02474}}.

\bibitem{Maldacena:1997re}
J.~M. Maldacena, \emph{{The Large N limit of superconformal field theories and
  supergravity}}, \href{http://dx.doi.org/10.1023/A:1026654312961,
  10.4310/ATMP.1998.v2.n2.a1}{\emph{Int. J. Theor. Phys.} {\bf 38} (1999)
  1113--1133}, [\href{https://arxiv.org/abs/hep-th/9711200}{{\tt
  hep-th/9711200}}].

\bibitem{Aharony:1999ti}
O.~Aharony, S.~S. Gubser, J.~M. Maldacena, H.~Ooguri and Y.~Oz, \emph{{Large N
  field theories, string theory and gravity}},
  \href{http://dx.doi.org/10.1016/S0370-1573(99)00083-6}{\emph{Phys. Rept.}
  {\bf 323} (2000) 183--386}, [\href{https://arxiv.org/abs/hep-th/9905111}{{\tt
  hep-th/9905111}}].

\bibitem{Gubser:1998bc}
S.~S. Gubser, I.~R. Klebanov and A.~M. Polyakov, \emph{{Gauge theory
  correlators from noncritical string theory}},
  \href{http://dx.doi.org/10.1016/S0370-2693(98)00377-3}{\emph{Phys. Lett. B}
  {\bf 428} (1998) 105--114}, [\href{https://arxiv.org/abs/hep-th/9802109}{{\tt
  hep-th/9802109}}].

\bibitem{Witten:1998qj}
E.~Witten, \emph{{Anti-de Sitter space and holography}},
  \href{http://dx.doi.org/10.4310/ATMP.1998.v2.n2.a2}{\emph{Adv. Theor. Math.
  Phys.} {\bf 2} (1998) 253--291},
  [\href{https://arxiv.org/abs/hep-th/9802150}{{\tt hep-th/9802150}}].

\bibitem{Pilch:2000ue}
K.~Pilch and N.~P. Warner, \emph{{N=2 supersymmetric RG flows and the IIB
  dilaton}}, \href{http://dx.doi.org/10.1016/S0550-3213(00)00656-8}{\emph{Nucl.
  Phys.} {\bf B594} (2001) 209--228},
  [\href{https://arxiv.org/abs/hep-th/0004063}{{\tt hep-th/0004063}}].

\bibitem{Buchel:2000cn}
A.~Buchel, A.~W. Peet and J.~Polchinski, \emph{{Gauge dual and noncommutative
  extension of an N=2 supergravity solution}},
  \href{http://dx.doi.org/10.1103/PhysRevD.63.044009}{\emph{Phys.Rev.} {\bf
  D63} (2001) 044009}, [\href{https://arxiv.org/abs/hep-th/0008076}{{\tt
  hep-th/0008076}}].

\bibitem{Evans:2000ct}
N.~J. Evans, C.~V. Johnson and M.~Petrini, \emph{{The Enhancon and N=2 gauge
  theory: Gravity RG flows}},
  \href{http://dx.doi.org/10.1088/1126-6708/2000/10/022}{\emph{JHEP} {\bf 10}
  (2000) 022}, [\href{https://arxiv.org/abs/hep-th/0008081}{{\tt
  hep-th/0008081}}].

\bibitem{Klebanov:2000hb}
I.~R. Klebanov and M.~J. Strassler, \emph{{Supergravity and a confining gauge
  theory: Duality cascades and chi SB resolution of naked singularities}},
  \href{http://dx.doi.org/10.1088/1126-6708/2000/08/052}{\emph{JHEP} {\bf 08}
  (2000) 052}, [\href{https://arxiv.org/abs/hep-th/0007191}{{\tt
  hep-th/0007191}}].

\bibitem{Herzog:2001xk}
C.~P. Herzog, I.~R. Klebanov and P.~Ouyang, \emph{{Remarks on the warped
  deformed conifold}},  in \emph{{Modern Trends in String Theory: 2nd Lisbon
  School on g Theory Superstrings Lisbon, Portugal, July 13-17, 2001}}, 2001.
\newblock \href{https://arxiv.org/abs/hep-th/0108101}{{\tt hep-th/0108101}}.

\bibitem{Maldacena:2000yy}
J.~M. Maldacena and C.~Nunez, \emph{{Towards the large N limit of pure N=1
  superYang-Mills}},
  \href{http://dx.doi.org/10.1103/PhysRevLett.86.588}{\emph{Phys. Rev. Lett.}
  {\bf 86} (2001) 588--591}, [\href{https://arxiv.org/abs/hep-th/0008001}{{\tt
  hep-th/0008001}}].

\bibitem{Candelas:1989js}
P.~Candelas and X.~C. de~la Ossa, \emph{{Comments on Conifolds}},
  \href{http://dx.doi.org/10.1016/0550-3213(90)90577-Z}{\emph{Nucl. Phys. B}
  {\bf 342} (1990) 246--268}.

\bibitem{Klebanov:1998hh}
I.~R. Klebanov and E.~Witten, \emph{{Superconformal field theory on
  three-branes at a Calabi-Yau singularity}},
  \href{http://dx.doi.org/10.1016/S0550-3213(98)00654-3}{\emph{Nucl. Phys.}
  {\bf B536} (1998) 199--218},
  [\href{https://arxiv.org/abs/hep-th/9807080}{{\tt hep-th/9807080}}].

\bibitem{Buchel:2018bzp}
A.~Buchel, \emph{{Klebanov-Strassler black hole}},
  \href{http://dx.doi.org/10.1007/JHEP01(2019)207}{\emph{JHEP} {\bf 01} (2019)
  207}, [\href{https://arxiv.org/abs/1809.08484}{{\tt 1809.08484}}].

\bibitem{Cassani:2010uw}
D.~Cassani, G.~Dall'Agata and A.~F. Faedo, \emph{{Type IIB supergravity on
  squashed Sasaki-Einstein manifolds}},
  \href{http://dx.doi.org/10.1007/JHEP05(2010)094}{\emph{JHEP} {\bf 05} (2010)
  094}, [\href{https://arxiv.org/abs/1003.4283}{{\tt 1003.4283}}].

\bibitem{Buchel:2000ch}
A.~Buchel, \emph{{Finite temperature resolution of the Klebanov-Tseytlin
  singularity}},
  \href{http://dx.doi.org/10.1016/S0550-3213(01)00051-7}{\emph{Nucl. Phys.}
  {\bf B600} (2001) 219--234},
  [\href{https://arxiv.org/abs/hep-th/0011146}{{\tt hep-th/0011146}}].

\bibitem{Buchel:2001gw}
A.~Buchel, C.~P. Herzog, I.~R. Klebanov, L.~A. Pando~Zayas and A.~A. Tseytlin,
  \emph{{Nonextremal gravity duals for fractional D-3 branes on the conifold}},
  \href{http://dx.doi.org/10.1088/1126-6708/2001/04/033}{\emph{JHEP} {\bf 04}
  (2001) 033}, [\href{https://arxiv.org/abs/hep-th/0102105}{{\tt
  hep-th/0102105}}].

\bibitem{Gubser:2001ri}
S.~S. Gubser, C.~P. Herzog, I.~R. Klebanov and A.~A. Tseytlin,
  \emph{{Restoration of chiral symmetry: A Supergravity perspective}},
  \href{http://dx.doi.org/10.1088/1126-6708/2001/05/028}{\emph{JHEP} {\bf 05}
  (2001) 028}, [\href{https://arxiv.org/abs/hep-th/0102172}{{\tt
  hep-th/0102172}}].

\bibitem{Aharony:2005zr}
O.~Aharony, A.~Buchel and A.~Yarom, \emph{{Holographic renormalization of
  cascading gauge theories}},
  \href{http://dx.doi.org/10.1103/PhysRevD.72.066003}{\emph{Phys. Rev.} {\bf
  D72} (2005) 066003}, [\href{https://arxiv.org/abs/hep-th/0506002}{{\tt
  hep-th/0506002}}].

\bibitem{Aharony:2007vg}
O.~Aharony, A.~Buchel and P.~Kerner, \emph{{The Black hole in the throat:
  Thermodynamics of strongly coupled cascading gauge theories}},
  \href{http://dx.doi.org/10.1103/PhysRevD.76.086005}{\emph{Phys. Rev.} {\bf
  D76} (2007) 086005}, [\href{https://arxiv.org/abs/0706.1768}{{\tt
  0706.1768}}].

\bibitem{Buchel:2009bh}
A.~Buchel, \emph{{Hydrodynamics of the cascading plasma}},
  \href{http://dx.doi.org/10.1016/j.nuclphysb.2009.06.001}{\emph{Nucl. Phys. B}
  {\bf 820} (2009) 385--416}, [\href{https://arxiv.org/abs/0903.3605}{{\tt
  0903.3605}}].

\bibitem{Buchel:2010wp}
A.~Buchel, \emph{{Chiral symmetry breaking in cascading gauge theory plasma}},
  \href{http://dx.doi.org/10.1016/j.nuclphysb.2011.01.031}{\emph{Nucl. Phys.}
  {\bf B847} (2011) 297--324}, [\href{https://arxiv.org/abs/1012.2404}{{\tt
  1012.2404}}].

\bibitem{Buchel:2021yay}
A.~Buchel, \emph{{A bestiary of black holes on the conifold with fluxes}},
  \href{http://dx.doi.org/10.1007/JHEP06(2021)102}{\emph{JHEP} {\bf 06} (2021)
  102}, [\href{https://arxiv.org/abs/2103.15188}{{\tt 2103.15188}}].

\bibitem{Buchel:2005nt}
A.~Buchel, \emph{{A Holographic perspective on Gubser-Mitra conjecture}},
  \href{http://dx.doi.org/10.1016/j.nuclphysb.2005.10.014}{\emph{Nucl. Phys.}
  {\bf B731} (2005) 109--124},
  [\href{https://arxiv.org/abs/hep-th/0507275}{{\tt hep-th/0507275}}].

\bibitem{Buchel:2005cv}
A.~Buchel, \emph{{Transport properties of cascading gauge theories}},
  \href{http://dx.doi.org/10.1103/PhysRevD.72.106002}{\emph{Phys. Rev. D} {\bf
  72} (2005) 106002}, [\href{https://arxiv.org/abs/hep-th/0509083}{{\tt
  hep-th/0509083}}].

\bibitem{Bena:2019sxm}
I.~Bena, A.~Buchel and S.~Lust, \emph{{Throat destabilization (for profit and
  for fun)}},  \href{https://arxiv.org/abs/1910.08094}{{\tt 1910.08094}}.

\bibitem{Buchel:2003ah}
A.~Buchel and J.~T. Liu, \emph{{Thermodynamics of the N=2* flow}},
  \href{http://dx.doi.org/10.1088/1126-6708/2003/11/031}{\emph{JHEP} {\bf 11}
  (2003) 031}, [\href{https://arxiv.org/abs/hep-th/0305064}{{\tt
  hep-th/0305064}}].

\bibitem{Gubser:2005ih}
S.~S. Gubser, \emph{{Phase transitions near black hole horizons}},
  \href{http://dx.doi.org/10.1088/0264-9381/22/23/013}{\emph{Class. Quant.
  Grav.} {\bf 22} (2005) 5121--5144},
  [\href{https://arxiv.org/abs/hep-th/0505189}{{\tt hep-th/0505189}}].

\bibitem{Hartnoll:2008kx}
S.~A. Hartnoll, C.~P. Herzog and G.~T. Horowitz, \emph{{Holographic
  Superconductors}},
  \href{http://dx.doi.org/10.1088/1126-6708/2008/12/015}{\emph{JHEP} {\bf 12}
  (2008) 015}, [\href{https://arxiv.org/abs/0810.1563}{{\tt 0810.1563}}].

\end{thebibliography}\endgroup

\end{document}